\documentclass[11pt,a4paper]{article}
\usepackage{jheppub}
\usepackage{amsmath,ascmac,amssymb}
\newcommand{\myeq}{\overset{\mathrm{AC}}{=}}
\newcommand{\scale}{\overset{\chi}{\sim }}
\title{High-Energy Expansion of
Two-Loop Massive Four-Point Diagrams}

\preprint{{TTP18-041}}

\author{Go Mishima}

\affiliation{Institute for Theoretical Particle Physics (TTP),
Karlsruhe Institute of Technology,\\
Wolfgang-Gaede-Stra{\ss}e 1,
76128 Karlsruhe, Germany}
\affiliation{Institute for Nuclear Physics (IKP),
Karlsruhe Institute of Technology,\\
Hermann-von-Helmholtz-Platz 1,
76344 Eggenstein-Leopoldshafen, Germany}
  
\emailAdd{go.mishima@kit.edu}


\abstract{
We apply the method of regions
to the massive two-loop integrals
appearing in the Higgs pair production cross section at the next-to-leading order,
in the high energy limit.
For the non-planar integrals,
a subtle problem arises 
because of the indefinite sign of
the second Symanzik polynomial.
We solve this problem by performing an analytic continuation 
of the Mandelstam variables
such that the second Symanzik polynomial has a definite sign.
Furthermore, we formulate the procedure of applying 
the method of regions systematically.
As a by-product of the analytic continuation of the Mandelstam variables,
we obtain crossing relations between integrals
in a simple and systematic way.
In our formulation, a concept of ``template integral" is introduced,
which represents and controls the contribution of each region.
All of the template integrals needed in the computation 
of the Higgs pair production at the next-to-leading order
are given explicitly.
We also develop techniques to solve Mellin-Barnes integrals,
and show them in detail.
Although most of the calculation is shown for the concrete example of
the Higgs pair production process,
the application to other similar processes is straightforward,
and we anticipate that our method can be useful also for other cases.
}

\begin{document}
\maketitle
\allowdisplaybreaks

\section{Introduction}
\label{ss:intro}

Perturbative quantum field theory
has been describing particle physics phenomena very well,
yet improving the perturbative series,
i.e., calculating Feynman integrals,
has been always challenging.
It becomes more and more important to include 
higher order corrections
to theoretical predictions
as particle physics experiments,
especially at the Large Hadron Collider (LHC),
become more and more precise over the years.
This means that a deeper understanding of higher order corrections
is required to obtain meaningful theoretical predictions.

One of the most interesting problems in this field
is higher order corrections to multi-scale processes.
The prime examples are 
the Higgs+jet,
Higgs pair,
and Higgs+$Z$ production cross sections at the LHC.
These $2\to 2$ processes involve
two kinematical variables, such as the center of mass energy and the transverse momentum,
and the masses of internal or external particles serve as additional scales.
Here, the word ``multi-scale" is used when there are more than three scales.
The bottlenecks of the calculation of multi-scale processes are
the integration-by-parts (IBP) reduction and
the evaluation of the resulting master integrals.
The subject of this paper is the second issue,
namely, the evaluation of multi-scale Feynman integrals.

Efforts to solve multi-scale Feynman integrals have persisted
over the years.
One of the milestones is the analytic computation of all the planar master integrals
contributing to the Higgs~$\to$~3~partons process at two-loops~\cite{Bonciani:2016qxi}.
Another milestone is the numeric evaluation of 
the Higgs+jet~\cite{Jones:2018hbb} and Higgs pair production~\cite{Borowka:2016ehy,Borowka:2016ypz} 
cross sections at two-loop level 
using the program \texttt{SecDec}~\cite{Borowka:2015mxa,Borowka:2017idc}.
An independent numerical evaluation of the Higgs pair production cross section
is given recently~\cite{Baglio:2018lrj}.
It is worth mentioning some recent analytic calculations of three-scale four-point two-loop diagrams;
the non-planar master integrals for $\mu e$ scattering ~\cite{Mastrolia:2017pfy,DiVita:2018nnh},
the planar double box integral relevant to top pair production~\cite{Adams:2018kez} and
the planar master integrals relevant to di-photon and di-jet production~\cite{Becchetti:2017abb}.
These works show that even three-scale problems are difficult to solve.
Recently, some of the non-planar master integrals for these processes
in the limit $m_H=0$ have been solved \cite{Xu:2018eos},
but there still remain unsolved master integrals.

It is a promising idea to reduce the number of scales
entering integrals
by expanding them in some small parameters.
For a summary of this topic, see Ref.~\cite{Smirnov:2002pj}.
In this direction, the large-$m_t$ expansion of
the Higgs+jet~\cite{Boughezal:2013uia,Chen:2014gva,Boughezal:2015dra,Boughezal:2015aha,Caola:2015wna,Chen:2016zka,Neumann:2016dny}
and Higgs pair production~\cite{deFlorian:2016uhr,deFlorian:2013uza,Grigo:2013rya,deFlorian:2013jea,Maltoni:2014eza,Grigo:2014jma,Grigo:2015dia,Degrassi:2016vss}
cross sections
is very well investigated.
However, it is not until recently that
expansions in other parameters
have been investigated.
Concerning the Higgs+jet production cross section,
the expansion in the small bottom quark mass $m_b\ll m_H$~\cite{Melnikov:2016qoc,Melnikov:2017pgf}
and in the small top quark and Higgs masses $m_t>m_H$~\cite{Kudashkin:2017skd}
are performed.
For the Higgs pair production cross section,
the expansion in small $m_t$
for the planar master integrals~\cite{Davies:2018ood}
and in small Higgs transverse momentum~\cite{Bonciani:2018omm}
are performed.
The rest of the master integrals of Higgs pair production 
in the small-$m_t$ expansion
are obtained in Ref.~\cite{Davies:2018qvx} together with the results of this paper.
Many of the non-planar master integrals are the same as,
or related to those of Ref.~\cite{Kudashkin:2017skd}
but we provide some new information needed for Higgs pair production.
Furthermore, the method used in this paper---the method of regions---is
completely different from the one in Ref.~\cite{Kudashkin:2017skd}
at all steps of the calculation,
so it provides a complementary understanding
of the massive non-planer integrals.
We would like to emphasize that 
the method of regions is
a generic and systematic procedure 
to expand integrals,
and thus the calculations shown in this paper
can be applied to other integrals 
in a straightforward way.

The concept of dividing the domain of integration variables
into several regions and expanding the integrand according
to hierarchies in each region
was introduced by Beneke and Smirnov~\cite{Beneke:1997zp}.
The method is now called ``expansion by regions"
or ``strategy of regions,"
and in this paper we call it the method of regions.
A mathematical proof of the method of regions 
for a general integral
is not yet known
although 
many successful applications 
have been reported.
In fact,
the author of the most up-to-date textbook
on this topic states in his book~\cite{Smirnov:2012gma}
\begin{quote}
\textit{
The strategy of expansion by regions still has
the status of experimental mathematics.
}
\end{quote}
In the cases of off-shell large-momentum expansion
and large-mass expansion,
a mathematical proof based on a graph-theoretical language
is known and it is called as ``expansion by subgraphs"~\cite{Smirnov:1990rz,Smirnov:2002pj}.
The procedure of expansion by subgraphs 
is implemented and can be performed in an automatic way~\cite{Harlander:1997zb,Seidensticker:1999bb}.
The large-$m_t$ expansion mentioned above belongs to this category.
A new proof of the method of regions was proposed by Jantzen~\cite{Jantzen:2011nz}
but its application is limited.
The purpose of this paper is to show
non-trivial examples where the method of regions
works well,
and our calculation shows the first application 
of the method to the high energy expansion 
of non-planar four-point integrals.

The remainder of the paper is organized as follows:
in Section~\ref{ss:gene},
we briefly summarize the method of regions.
In Section~\ref{ss:setup}
we introduce conventions, ideas, and techniques,
which will be used in the following sections.
In Section~\ref{ss:one}, \ref{ss:twoPL}  and~\ref{ss:two},
we apply the method of regions to the one-loop box diagram,
the two-loop planar massive diagrams,
and
the two-loop non-planar massive diagrams,
respectively.

%%%%%%%%%%%%%%%%%%%%%%%%%%%%%%%%%%%%%%%%%%%%%%%%%%%%%%%
\section{General Idea of the Method of Regions}
\label{ss:gene}

The procedure 
of the method of regions is the following
~\cite{Beneke:1997zp,Smirnov:2002pj,
Jantzen:2011nz,Smirnov:2012gma,
Semenova:2018cwy}:
\begin{itemize}
\item Step 1: Assign a hierarchy to the dimensionful parameters.
\item Step 2: Reveal the relevant scaling of the integration variable.
\item Step 3: For each region, expand the integrand according to its scaling.
\item Step 4: Integrate. Scaleless integrals such as $\int _0^\infty dx~x^a$ are set to zero.
\item Step 5: Sum over the contributions from all the relevant regions.
\end{itemize}
The Step~2 is the crucial part of the method of regions,
and an algorithm to reveal such scalings for a general integral 
is established 
based on the analysis of the convex hull~\cite{Pak:2010pt,Jantzen:2012mw}.
One can use the algorithm, 
implemented in the \texttt{Mathematica} package \texttt{asy2.1.m}~\cite{Jantzen:2012mw}.
Although it is not proved that the algorithm works correctly for all the cases,
no counterexample is known so far.
Recently,
a new idea to reveal relevant scalings is proposed
based on the technique of power geometry,
which is implemented in the \texttt{Mathematica} 
package \texttt{ASPIRE}~\cite{Ananthanarayan:2018tog}.
In this paper we use \texttt{asy2.1.m}.

The practical bottleneck is Step~4,
since the integration tends to be complicated
even after the expansion
if the original integral is very complicated.
This is one of the reasons why
testing the method of regions is difficult.

The method of regions was first applied to the momentum representation
of the Feynman integrals,
so the ``regions" mean some domains of the loop momenta.
Later, it was found that parametric representations
such as the Feynman representation and the alpha representation
are more convenient to apply the method~\cite{Smirnov:1999bza}.
Recently, it was proposed in Ref.~\cite{Semenova:2018cwy}
to use yet another parametric representation,
Lee-Pomeransky representation~\cite{Lee:2013hzt},
to apply the method of regions.
For all the representations mentioned above,
one has to follow Step 1 to 5
for practical calculation.

%%%%%%%%%%%%%%%%%%%%%%%%%%%%%%%%%%%%%%%%%%%%%%%%%%%%%%%
\section{Notation and Technical Tools}
\label{ss:setup}

%%%%%%%%%%%%%%%%%%%%%%%%%%%%%%%%%%%%%%%%%%%%%%%%%%%%%%%
\subsection{Conventions}
\label{ss:conv}

We distinguish the exact equal sign $``="$
and the equal sign under a certain analytic continuation.
For this purpose, we introduce a sign~$``\myeq "$
and use it as, e.g.,
\begin{align}
\log (z+i0)\myeq \log (-z-i0)+i\pi
\,,
\label{ac}
\end{align}
where $i0$ represents an infinitesimal positive imaginary number.
We interpret $\log(z)$ as the principal value of the complex logarithm
whose imaginary part lies in the interval $(-\pi,\pi ]$.
Both the left-hand side and the right-hand side
of Eq.~\eqref{ac}
are well-defined in the entire domain of $z$,
but the equality is valid only in the upper half plane of $z$.
This is how analytic continuation is 
%usually 
performed,
and that is why we add ``AC" to
the normal equal sign in Eq.~\eqref{ac}.
The equality of a series expansion like
\begin{align}
\frac{1}{1-m/M}=\sum_{n=0}^\infty \left( \frac{m}{M} \right) ^n
\label{series}
\end{align}
is in principle also regarded as
an analytic continuation.
However, when a hierarchy like $m\ll M$ is explicitly stated in the text,
we use normal equal sign.

We use a simplified expression
of the Landau $\mathcal{O}$ notation 
for more than one variable as
\begin{align}
X+
\mathcal{O}\left( (m_H^2)^{n_H},(m_t^2)^{n_t}, \epsilon ^n\right)
\equiv 
X+
\mathcal{O}\left( (m_H^2)^{n_H}\right)
+\mathcal{O}\left((m_t^2)^{n_t}\right)
+\mathcal{O}\left( \epsilon ^n\right)
\,.
\end{align}
The Euler--Mascheroni constant is denoted as $\gamma_E$.

We use the alpha representation
to calculate Feynman integrals.
The integration measure and the analytic regularization parameters
are defined as
\begin{align}
\int\! \mathfrak{D}^n \alpha^\delta
\equiv 
\prod_{i=1}^n \left( \int_0^\infty 
\frac{\mathrm{d}\alpha _i ~\alpha_i^{\delta_j}}{\Gamma (1+\delta_j)} \right)
\,.
\label{one-da}
\end{align}
The analytic regularization parameters $\delta_j$ play 
one essential role and three secondary roles:
%%%%%%%%%%%%%%%%%%%%%%%%%%
\renewcommand{\theenumi}{(\roman{enumi})}
\begin{enumerate}
\item The essential role
is to regularize the contribution of individual regions
which are divergent if we naively expand in $\alpha_i$.
This means that individual contributions are
regulator dependent,
and the dependence on $\delta_j$ cancel after
we sum all the contributions
and take the limits $\delta_j\to0$.
In taking the limit,
it is necessary to specify the order 
because some of them do not commute.
We express the sequence of limits as
\begin{align}
\lim_{\epsilon,\delta_n,...,\delta_2,\delta_1\to0}
X
\equiv 
\lim_{\epsilon\to0}
\lim_{\delta_n\to0}
\cdots
\lim_{\delta_2\to0}
\lim_{\delta_1\to0}
X\,.
\end{align}
\item We use $\delta_j$ to regularize the Mellin-Barnes integral.
[See the text below Eq.~\eqref{mb2}.]
\item By shifting $\delta_j\to\delta_j+1$,
we can express polynomials of $\alpha_i$ in the integrand.
For example, when $n=2$,
\begin{align}
\int\! \mathfrak{D}^2 \alpha^\delta
\left( \alpha_1^2+\alpha_1\alpha_2\right)
=
\left.
\int\! \mathfrak{D}^2 \alpha^\delta
\right|_{\delta_1\to \delta_1+2}
+
\left.
\int\! \mathfrak{D}^2 \alpha^\delta
\right|_{\delta_1\to \delta_1+1,\delta_2\to\delta_2+1}
\label{del-shift}
\end{align}
This property is usually used to express the integrals
with higher powers of propagators.
\item We use the property of Eq.~\eqref{del-shift} 
to express the higher order terms.
[See the text below Eq.~\eqref{one-r1-3}.]
\end{enumerate}
%%%%%%%%%%%%%%%%%%%%%%%%%%

The sum of the variables will be expressed as
\begin{align}
\alpha_{i_1...i_n}
\equiv 
\alpha_{i_1}+\cdots +\alpha_{i_n}
,\qquad
\delta_{i_1...i_n \overline{i_{n+1}}...i_{n'}}
\equiv 
\delta_{i_1}+\cdots +\delta_{i_n}
-\delta_{i_{n+1}}+\cdots+\delta_{i_{n'}}
\,.
\end{align}
The bar on an index indicates that
the variable corresponding to the index
is subtracted instead of added.
Sometimes $\epsilon$ and $\delta_j$ are treated in the same way,
and in those cases we express $\epsilon$ as $\delta_0$.
For example, $\delta_{001\bar2}=2\epsilon+\delta_1-\delta_2$.

Also, we introduce the following compact notation for the product of $\Gamma$-functions
\begin{align}
\Gamma\left[ x_1,\dots,x_n \right]
\equiv 
\prod_{i=1}^n \Gamma (x_i)
\,.
\label{gamma}
\end{align}

In Step~3 of Section~\ref{ss:gene},
we expand the integrand of Feynman integrals
in terms of soft parameters.
In order to control the expansion
in a systematic way,
we introduce an auxiliary soft-scaling parameter~$\chi$.
For example,
assume that we have four variables
$\alpha_1,...,\alpha_4$ whose scalings are
\begin{align}
\alpha_1\sim m,\quad 
\alpha_2\sim M,\quad 
\alpha_3\sim m,\quad 
\alpha_4\sim M, 
\label{chi-scale}
\end{align}
where $m\sim \chi$ is a soft parameter 
and $M\sim 1$ is a hard parameter.
In this case,
we apply a substitution 
\begin{align}
m\to \chi m,\quad
M\to M,\quad
\alpha_1\to \chi\alpha_1,\quad
\alpha_2\to \alpha_2,\quad
\alpha_3\to \chi\alpha_3,\quad
\alpha_4\to \alpha_4,
\end{align}
to the integrand and
expand in $\chi$.
After that,
we can set $\chi =1$.
In this paper we denote the scalings~\eqref{chi-scale} as
\begin{align}
(\alpha_1,\alpha_2,\alpha_3,\alpha_4)
\scale (1,0,1,0)
\end{align}
or simply $(1,0,1,0)$.

%%%%%%%%%%%%%%%%%%%%%%%%%%%%%%%%%%%%%%%%%%%%%%%%%%%%%%%
\subsection{Kinematics and High Energy Expansion}
\label{ss:kinematics}

\begin{figure}
\centering
%\begin{tikzpicture}[scale=0.25]
%\draw [ultra thick] (4,4)--(4,-4)--(-4,-4)--(-4,4)--(4,4);
%\draw [ultra thick] (4,4)--(5,5);
%\draw [ultra thick, <-] (5,5)--(7,7);
%\draw [ultra thick] (4,-4)--(5,-5);
%\draw [ultra thick,<-] (5,-5)--(7,-7);
%\draw [ultra thick] (-4,4)--(-5,5);
%\draw [ultra thick,<-] (-5,5)--(-7,7);
%\draw [ultra thick] (-4,-4)--(-5,-5);
%\draw [ultra thick,<-] (-5,-5)--(-7,-7);
%\draw (-4,6) node [above] {$q_1$};
%\draw (-4,-6) node [below] {$q_2$};
%\draw (4,6) node [above] {$q_3$};
%\draw (4,-6) node [below] {$q_4$};
%\end{tikzpicture}
\includegraphics[width=0.3\textwidth]{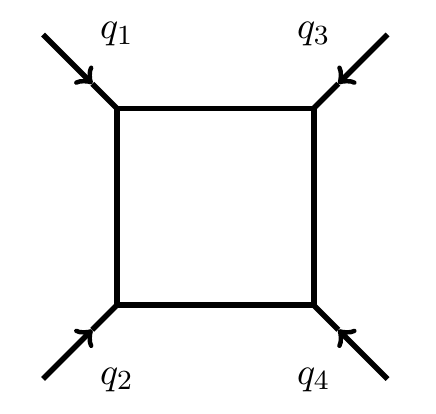}
\caption{The convention of the external momenta.}
\label{fig:mom}
\end{figure}

The assignment of the external momenta $q_1,...,q_4$
is illustrated in Fig.~\ref{fig:mom}.
We consider the 2 to 2 process 
but define all the external momenta as incoming.
In addition to the usual Mandelstam variables $s, t, u,$
(which we call physical Mandelstam variables),
we introduce $S, T, U$ as
\begin{align}
S=-s=-(q_1+q_2)^2,\qquad
T=-t=-(q_1+q_3)^2,\qquad
U=-u=-(q_2+q_3)^2
\,.
\label{stu}
\end{align}
In our calculation we assume that $S,T,$ and $U$ are positive and thus
call them positive Mandelstam variables.
Sometimes two of the three Mandelstam variables
are sufficient to express four-point functions,
and indeed for planar integrals we do not use $U$
[See Section~\ref{ss:twoPL}].
However for non-planar integrals,
all of $S,T,U$ are required 
to make the second Symanzik polynomial positive
[See Subsection~\ref{ss:two-scale}].

In this paper, 
we consider the master integrals 
of Higgs pair production at next-to-leading order,
where the loops are induced by the top quark.
Therefore there are two mass scales, the Higgs mass $m_H$ and the top quark mass $m_t$.
We consider the high energy limit where
the following hierarchy is satisfied 
\begin{align}
m_H^2<m_t^2\ll S,T,U
\,.
\label{hie0}
\end{align}
The high energy expansion in this case is two-fold.
First, we treat $m_H^2$ as the soft parameter
and $m_t^2,S,T,U$ as the hard parameters.
Afterwards, we treat $m_t^2$ as the soft parameter
and $S,T,U$ as the hard parameters.

In the first expansion, i.e. the $m_H$-expansion,
$m_H$ enters the integrals through 
the on-shell condition of the external momenta
\begin{align}
  q_1^2=0,\quad
  q_2^2=0,\quad
  q_3^2=m_H^2,\quad
  q_4^2=m_H^2
  \,.
  \label{os1}
\end{align}
In terms of $\chi$,
the scaling we impose here is
\begin{align}
m_H^2\sim \chi,\quad
m_t^2\sim 1,\quad
S\sim 1,\quad
T\sim 1,\quad
U\sim 1\,,
\label{sca1}
\end{align}
and the resulting series expression of an integral $I$
is expressed symbolically as\footnote{
In general,
the coefficients $c_{n_H}$ depend on
$\log (m_H^2)$,
but in our case not.
}
\begin{align}
I(S,T,U,m_t^2,m_H^2)
=\sum_{n_H} (m_H^2)^{n_H} c_{n_H} (S,T,U,m_t^2)
\,.
\label{symbol1}
\end{align}
A note of caution should be made regarding
the dependence on $S,T,U$.
Since the physical Mandelstam variables satisfy
the relation $s+t+u=2m_H^2$,
a similar relation should also hold for the positive Mandelstam variables.
As a result, functions expressed in terms of $S,T,U$
are not unique.
However this is not a problem,
because 
at the end of the calculation,
we express the result in a unique way.
[See the text below Eq.~\eqref{anaconS}.]
We use the linear dependence of $S,T,U$
to make the second Symanzik polynomial positive definite.
[See Subsection~\ref{ss:two-scale}.]
After the $m_H$-expansion,
the external legs becomes massless legs.

In the second expansion, i.e. the $m_t$-expansion,
the on-shell condition of the external momenta becomes 
\begin{align}
  q_1^2=0,\quad
  q_2^2=0,\quad
  q_3^2=0,\quad
  q_4^2=0
  \,.
  \label{os2}
\end{align}
In terms of $\chi$,
the scaling we impose here is
\begin{align}
m_t^2\sim \chi,\quad
S\sim 1,\quad
T\sim 1,\quad
U\sim 1\,,
\label{sca2}
\end{align}
and the resulting series for an integral $I$ 
is expressed symbolically as
\begin{align}
I(S,T,U,m_t^2,m_H^2)
=\sum_{n_H} (m_H^2)^{n_H} 
\sum_{n_t} (m_t^2)^{n_t} c_{n_H,n_t} (S,T,U,\log (m_t^2))
\,.
\label{symbol2}
\end{align}

The result should be expressed
in a way suitable for the evaluation with the physical kinematics.
In order to achieve that,
the analytic continuation 
\begin{align}
S\myeq e^{-i\pi +i0} (s+i0),\quad
T\myeq sv,\quad
U\myeq s(1-v),
\label{anaconS}
\end{align}
is applied,
where $i0$ represents an infinitesimal positive imaginary number,
(note that the massless on-shell condition~\eqref{os2}
is adopted), 
and
$0\leq v\leq 1$
in the physical kinematics.
After expressing the result in terms of $s$ and $v$,
the expression is unique.
The analytic continuation of $T,U$ is trivial
since their signs are consistent with those of the physical kinematics.
The results are expressed in terms of
harmonic polylogarithms~(HPL)~\cite{Remiddi:1999ew}
and we introduce an abbreviation for HPL as
\begin{align}
h_0=H(0;v),\quad
h_1=H(1;v),\quad
h_2=H(2;v),\quad
h_{2,1}=H(2,1;v),
\end{align}
and so on.
The argument of the HPL is always chosen to be $v$.
For example,
\begin{align}
H\left(3;-\frac{u}{s}\right)
=
H\left(3;1-v\right)
\myeq
-h_{2,1}+h_2h_1-\frac{1}{2} h_1^2h_0-\frac{\pi^2}{6}h_1+\zeta_3
\,.
\label{hpl3}
\end{align}
We use the \texttt{Mathematica} package
\texttt{HPL.m}~\cite{Maitre:2005uu,Maitre:2007kp} in dealing with HPL.

\begin{figure}
\centering
%\begin{tikzpicture}[scale=0.3]
%\draw (-3,5) node [left] {$F(S,T,U)$};
%\draw (3,5) node [right] {$F(U,T,S)$};
%\draw (-3,-5) node [left] {$f(s,v)$};
%\draw (3,-5) node [right] {$\tilde f(s,v)$};
%\draw [->] (-3,5)--(3,5);
%\draw (0,5) node [above] {crossing};
%\draw [->] (-3,-5)--(3,-5);
%\draw (0,-5) node [below] {crossing};
%\draw (0,-6.2) node [below] {(analytic continuation required)};
%\draw [->] (-5,3)--(-5,-3);
%\draw (-5,0) node [left] {analytic continuation};
%\draw [->] (5,3)--(5,-3);
%\draw (5,0) node [right] {analytic continuation};
%\draw [->] (-2,4) arc [start angle=90, end angle=0, radius=6];
%\draw (3,3) node {(a)};
%\draw [->] (-2,-4)--(2,-4);
%\draw (0,-4) node [above] {(b)};
%\end{tikzpicture}
\includegraphics[width=0.7\textwidth]{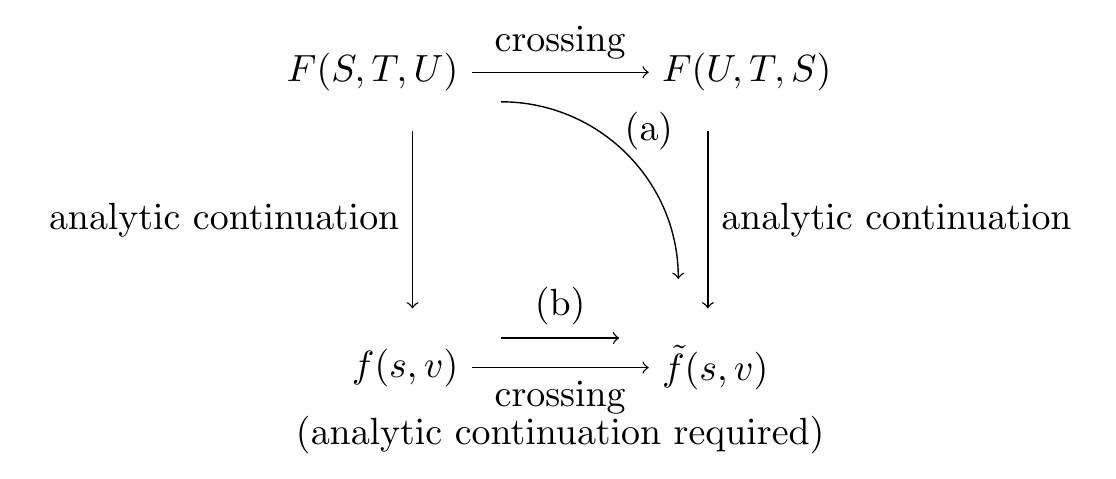}
\caption{The commutative diagram
of the crossing and the analytic continuation.
Note that the crossing at upper level is a literal replacement
of $S$ and $U$ whereas
the crossing at bottom level changes
the function in a nontrivial way.
(a) Our approach. 
(b) Conventional approach.}
\label{fig:com}
\end{figure}

When diagrams of the same topology but with a different assignment
of external momenta are considered,
it is convenient to relate them
by applying replacements of the external momenta,
which means replacements of the Mandelstam variables.
We call these replacements ``crossing relations".
This subject is already well-established in the case of HPL~\cite{Anastasiou:2000mf},
but we propose a simpler way to obtain the crossing relations.
In Fig.~\ref{fig:com},
the commutative diagram of the crossing and analytic continuation is given.
Usually an integral is given as the bottom-left expression,
where the result is expressed in terms of the physical kinematic variables.
On the other hand, we proceed the crossing in the upper expression,
where the result is expressed in terms of the positive Mandelstam variables.
The analytic continuation in the upper expression
is the simple replacement of the positive Mandelstam variables,
whereas the analytic continuation of the bottom expression 
requires the precise knowledge of the branch cuts.
We take the approach (a) of Fig.~\ref{fig:com}
because it is easy to implement in programs,
and crosscheck the result using approach (b).
One can interpret the simplification by introducing the positive Mandelstam variables
as the resolution of singularities by increasing the dimension,
or in another words,
we lose some information when we map the three-variable function $F$
into the two-variable function $f$.

We would like to emphasize that
the method to obtain the crossing relations explained above
is a by-product of introducing the positive Mandelstam variables.
The most important point in introducing the positive Mandelstam variables
is that it makes the Symanzik polynomials positive and
thus allows us to apply the method of regions safely.
[See Subsection~\ref{ss:two-scale}.]

%%%%%%%%%%%%%%%%%%%%%%%%%%%%%%%%%%%%%%%%%%%%%%%%%%%%%%%
\section{A First Example: One-Loop Box Diagram}
\label{ss:one}

\begin{figure}
\centering
%\begin{tikzpicture}[scale=0.5]
%  \draw [ultra thick] (0,0)--(4,0)--(4,4)--(0,4)--(0,0);
%  \draw [dotted, ultra thick] (0,0)--(-1,-1);
%  \draw [dotted, ultra thick] (4,0)--(5,-1);
%  \draw [dotted, ultra thick] (4,4)--(5,5);
%  \draw [dotted, ultra thick] (0,4)--(-1,5);
%  \draw (-1,5) node [above] {$q_1^2=0$};
%  \draw (-1,-1) node [below] {$q_2^2=0$};
%  \draw (5,5) node [above] {$q_3^2=m_H^2$};
%  \draw (5,-1) node [below] {$q_4^2=m_H^2$};
%  \draw (2,4) node [above] {$\ell$};
%  \draw (2,0) node [below] {$\ell+q_1+q_2$};
%  \draw (0,2) node [left] {$\ell+q_1$};
%  \draw (4,2) node [right] {$\ell-q_3$};
%\end{tikzpicture}
\includegraphics[width=0.35\textwidth]{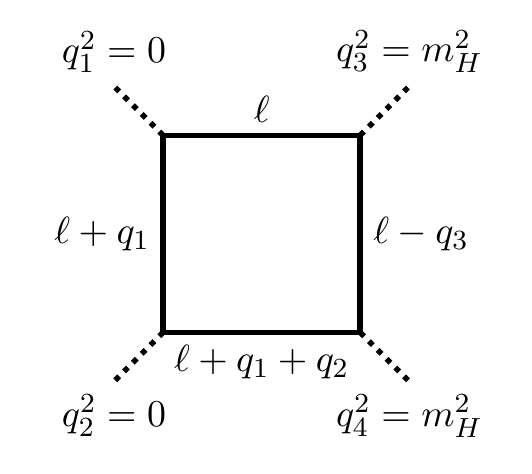}
\caption{The one-loop box diagram considered 
in Section~\ref{ss:one}}
\label{fig:one}
\end{figure}

We consider the massive one-loop Feynman integral family
\begin{align}
J_{a_1,a_2,a_3,a_4}=
  \int\!\!
  \frac{d^d\ell}{i \pi^{d/2}}
  \frac{1}
  {\left(m_t^2-\ell^2\right)^{a_1}
  \left( m_t^2-(\ell+q_1)^2 \right)^{a_2}
    \left( m_t^2-(\ell+q_1+q_2)^2 \right)^{a_3}
  \left( m_t^2-(\ell-q_3)^2 \right)^{a_4}}
  \,,
  \label{one1}
\end{align}
where the infinitesimal negative imaginary part of each denominator is implicit.
The external momenta $q_i$ satisfy the on-shell conditions~\eqref{os1}.
We consider the box diagram shown in Fig.~\ref{fig:one},
$J_{1,1,1,1}$
and its alpha representation is
\begin{align}
J_{1,1,1,1}
\myeq
I&=\int\! \mathfrak{D}^4\alpha^\delta~
  \mathcal{U}^{-d/2}
  e^{ -\mathcal{F}/\mathcal{U} }\,,
  \label{one-alpha}
\end{align}
where the first Symanzik polynomial $\mathcal{U}$
and the second Symanzik polynomial $\mathcal{F}$
are given by
\begin{align}
\mathcal{U}=\alpha_{1234},\qquad 
\mathcal{F}=m_t^2\alpha_{1234}~\mathcal{U}+S\alpha_1\alpha_3+T\alpha_2\alpha_4-m_H^2\alpha_{13}\alpha_4
\,.
\label{one-uf}
\end{align}
We make clear the analytic continuation in Eq.~\eqref{one-alpha}
because the right hand side is regularized by $\delta_j$
whereas Eq.~\eqref{one1} is explicitly $\delta_j$-independent.
Also, we assume $m_H^2<0$ in order to ensure the convergence of the integral
and perform the analytic continuation of the result to $m_H^2>0$ at the end,
which turns out to be trivial.

%%%%%%%%%%%%%%%%%%%%%%%%%%%%%%%%%%%%%%%%%%%%%%%%%%%%%%%
\subsection{Expansion in the Higgs Mass}
\label{ss:one-higgs}

We first expand in $m_H$.
By using \texttt{asy2.1.m},
we find that
there is only one relevant scaling
\begin{align}
(\alpha_1,\alpha_2,\alpha_3,\alpha_4)
\scale (0,0,0,0)
\,.
\end{align}
The expansion corresponding to this scaling 
is actually just the Taylor expansion of the integrand
in $m_H^2$,
and the original integral can be written as
\begin{align}
I&=\int\! \mathfrak{D}^4\alpha^\delta~
  \mathcal{U}^{-d/2}
  \left[
  e^{ -\mathcal{F}/\mathcal{U} } \Big|_{(m_H^2)=0}
  +(m_H^2) \frac{\partial (e^{ -\mathcal{F}/\mathcal{U} })}{\partial (m_H^2)}  \Big|_{(m_H^2)=0}
  +\cdots
  \right]\,.
  \label{one-mh-exp}
\end{align}
In particular, the leading term is identical to the box diagram with completely massless external legs.

The fact that 
the expansion in $m_H^2$
and the integration commute
is reasonable
because $m_H$ in the denominator of the integrand
(appearing through $q_3^2=m_H^2$)
is always accompanied by $m_t$
which regulates the integral
and the limit $m_H\to 0$ always exists.

In collider physics,
it is common to use $s$ and 
the transverse momentum $p_T$ as the kinematic variables,
where $t$ and $p_T$ are related as
\begin{align}
p_T^2=\frac{ut-m_H^4}{s}
\,,
\label{mh-3b}
\end{align}
or equivalently\footnote{
When we solve Eq.~\eqref{mh-3b} in $t$
using the relation $s+t+u=2m_H^2$,
there are two solutions.
The other solution has ``+" in front of the square root in Eq.~\eqref{mh-3},
and it corresponds to $u$ in this case.
Note that the amplitude is symmetric under $t\leftrightarrow u$,
so we could choose the sign the other way around.
We choose the sign such that 
$t= -p_T^2+\mathcal{O}(p_T^2/s,m_H^2/s)$.
}
\begin{align}
t=-\frac{1}{2}
\left[
s-2m_H^2
- \sqrt{s(s-4m_H^2-4p_T^2)}
\right]
\,.
\label{mh-3}
\end{align}
Since these relations are $m_H$-dependent,
one should be careful when expanding in $m_H$.
In order to clarify the point,
let us consider two expressions
\begin{align}
f_1(t,m_H^2)=f_2(p_T^2,m_H^2)
\label{mh-4}
\end{align}
which are related by Eq.~\eqref{mh-3}.
We would like to analyze the cross section for fixed $p_T$ but not $t$,
so let us consider the $m_H$-expansion of $f_2(p_T^2,m_H^2)$:
\begin{align}
f_2(p_T^2,m_H^2)
=f_2(p_T^2,0)
+(m_H^2) \left.\frac{\partial f_2}{\partial (m_H^2)}\right|_{m_H=0}
+\frac{(m_H^2)^2}{2} \left.\frac{\partial ^2f_2}{\partial (m_H^2)^2}\right|_{m_H=0}
+\mathcal{O}((m_H^2)^3)
\,.
\label{mh-5}
\end{align}
On the other hand, the kinematic variable appearing in the Feynman integral
is $t$
and the natural representation is $f_1(t,m_H^2)$.
Thus, we express the ingredients of
Eq.~\eqref{mh-5} in terms of $f_1(t,m_H^2)$ as
\begin{align}
f_2(p_T^2,0)
=&f_1(t_0,0)
\label{mh-6}
\\
\left.\frac{\partial f_2}{\partial (m_H^2)}\right|_{m_H=0}
=&
\left.\frac{\partial t}{\partial (m_H^2)}\right|_{m_H=0}
\left.\frac{\partial f_1}{\partial t}\right|_{t=t_0,m_H=0}
+\left.\frac{\partial f_1}{\partial (m_H^2)}\right|_{t=t_0,m_H=0}
\label{mh-7}
\\
\left.\frac{\partial ^2f_2}{\partial (m_H^2)^2}\right|_{m_H=0}
=&
\left.\frac{\partial ^2t}{\partial (m_H^2)^2}\right|_{m_H=0}
\left.\frac{\partial f_1}{\partial t}\right|_{t=t_0,m_H=0}
+\left[ \left.\frac{\partial t}{\partial (m_H^2)}\right|_{m_H=0} \right]^2
\left.\frac{\partial ^2f_1}{\partial t^2}\right|_{t=t_0,m_H=0}
\nonumber\\
&
+2\left.\frac{\partial t}{\partial (m_H^2)}\right|_{m_H=0}
\left.\frac{\partial ^2f_1}{\partial t\partial (m_H^2)}\right|_{t=t_0,m_H=0}
+\left.\frac{\partial ^2f_1}{\partial (m_H^2)^2}\right|_{t=t_0,m_H=0}
\label{mh-8}
\end{align}
where $t_0=t|_{m_H=0}$.
Apparently,
Eq.~\eqref{mh-5}
becomes complicated 
when Eqs.~\eqref{mh-6},~\eqref{mh-7},~\eqref{mh-8}
are substituted.
However, taking into account the $m_H$-expansion of $f_1(t,0)$
\begin{align}
f_1(t,0)=f_1(t_0,0)+
(m_H)^2
\left.\frac{\partial t}{\partial (m_H^2)}\right|_{m_H=0}
\left.\frac{\partial f_1}{\partial t}\right|_{t=t_0,m_H=0}
+\cdots
\,,
\end{align}
the expression becomes simpler
\begin{align}
f_2(p_T^2,m_H^2)
=f_1(t,0)
+(m_H^2) \left.\frac{\partial f_1}{\partial (m_H^2)}\right|_{m_H=0}
+\frac{(m_H^2)^2}{2} \left.\frac{\partial ^2f_1}{\partial (m_H^2)^2}\right|_{m_H=0}
+\mathcal{O}((m_H^2)^3)
\,,
\end{align}
where the $m_H$-dependent $t$~\eqref{mh-3} is used.
In general,
for a given order of $m_H^2$,
the difference between the strict $m_H$-expansion 
of $f_2(p_T^2,m_H^2)$ for fixed $p_T$
and the $m_H$-expansion of $f_1(t,m_H^2)$
for the $m_H$-dependent $t$
is higher order in $m_H^2$.

The conclusions of this subsection are the following:
\begin{enumerate}
\item As the result of the $m_H$-expansion,
the integrals reduce to integrals with massless external legs.
\item The $m_H$-expansion for fixed $p_T$
is obtained by
the $m_H$-expansion with fixed~$t$,
keeping the $m_H$-dependence of $t$.
\end{enumerate}

%%%%%%%%%%%%%%%%%%%%%%%%%%%%%%%%%%%%%%%%%%%%%%%%%%%%%%%
\subsection{Expansion in the Top Quark Mass}
\label{ss:one-top}

After the expansion in $m_H$,
we have integrals which depend on $m_t$, $S,$ and $T$.
We now consider the expansion in $m_t$ assuming the hierarchy~\eqref{sca2}.
The integral of interest is 
\begin{align}
\int\! \mathfrak{D}^4\alpha^\delta~
  \mathcal{U}^{-d/2}
  e^{ -\mathcal{F}/\mathcal{U} },
  \label{one-alpha2}
\end{align}
with
\begin{align}
\mathcal{U}=\alpha_{1234},\qquad 
\mathcal{F}=m_t^2\alpha_{1234}\mathcal{U}+S\alpha_1\alpha_3+T\alpha_2\alpha_4
\,.
\label{one-uf2}
\end{align}
Here we use the positive Mandelstam variables, $S,T$,
to make all the terms in $\mathcal{F}$ positive.
Otherwise, hard terms could cancel 
and result in a soft term,
which breaks the method of regions.
The use of positive Mandelstam variables in Eq.~\eqref{one-uf2}
is conceptually not new,
since it corresponds to the integral in the $u$-channel
where $s<0,t<0$ and $u=-s-t>0$.
The absence of a negative term in the $u$-channel
is reasonable because there is no physical cut in those kinematics.

By using the package \texttt{asy2.1.m}~\cite{Jantzen:2012mw},
we reveal five relevant scalings:\footnote{
More precisely,
\texttt{asy2.1.m}
reveals the scalings 
which lead to homogeneous and non-scaleless
integrals.
}
\begin{align}
\underbrace{(0,0,0,0)}_{1},~
\underbrace{(0,0,1,1)}_{2},~
\underbrace{(0,1,1,0)}_{3},~
\underbrace{(1,0,0,1)}_{4},~
\underbrace{(1,1,0,0)}_{5}\,.
\label{one-scale}
\end{align}
The scalings of regions 2 to 5 reflect
the symmetries of the integral,
$\alpha_1\leftrightarrow\alpha_3$
and $\alpha_2\leftrightarrow\alpha_4$.
Eq.~\eqref{one-alpha2} is thus expressed as the sum
of the contributions from these five regions:
$\sum_{i=1}^5 I^{(i)}$.

%%%%%%%%%%%%%%%%%%%%%%%%%%%%%%%%%%%%%%%%%%%%%%%%%%%%%%%
\subsubsection*{Region 1 (all-hard region)}
The region where all the alpha variables scale as $\chi^0$,
i.e., $(\alpha_1,\alpha_2,\alpha_3,\alpha_4)\scale (0,0,0,0)$
is special,
and we call this region the ``all-hard region".
We can make several general statements about this region
within our high-energy expansion:
\begin{enumerate}
\renewcommand{\theenumi}{(\alph{enumi})}
\item Every integral has one all-hard region.
\item There is only one soft parameter in the all-hard region, which is $m_t$.
\item The contribution from the all-hard region can be expressed as the massless integral of the original topology.
	In particular, the leading order term is obtained by substituting $m_t=0$ into the original integral.
\item The leading order contribution of the all-hard region is $\mathcal{O}(\chi ^0)$.
\item The contribution from the all-hard region has no singularities in $\delta_j$.
\end{enumerate}
Because of these properties,
the contribution from the all-hard region
can be calculated in two ways.
The first is the procedure universal for any region,
and the second is to use the momentum representation.
We show them in order.

First, we show the universal procedure.
By expanding Eq.~\eqref{one-alpha2} in terms of $\chi m_t^2$,
we obtain the contribution of this region as
\begin{align}
I^{(1)}&=\int\! \mathfrak{D}^4\alpha^\delta~
  {\mathcal{U}^{(1)}}^{-d/2}
  e^{ -\mathcal{F}^{(1)}/\mathcal{U}^{(1)} }
	\left[
	1-\chi m_t^2\alpha_{1234}
	+\frac{\chi^2}{2} \left( m_t^2\alpha_{1234} \right) ^2
	\right]
	+\mathcal{O}(\chi^3)
\label{one-r1-1}
\,,
\end{align}
where $\mathcal{U}^{(1)}=\alpha_{1234}$ and $\mathcal{F}^{(1)}=S\alpha_1\alpha_3+T\alpha_2\alpha_4$.
As stated above,
the leading order term is the massless box integral
and we name it $\mathcal{T}^{(1)}_{\delta_1,\delta_2,\delta_3,\delta_4,\epsilon}$
for later use:
\begin{align}
&\mathcal{T}^{(1)}_{\delta_1,\delta_2,\delta_3,\delta_4,\epsilon}
=\int\! \mathfrak{D}^4\alpha^\delta~
  {\mathcal{U}^{(1)}}^{-d/2}
  e^{ -\mathcal{F}^{(1)}/\mathcal{U}^{(1)} }
\label{one-r1-2}
\\
&=\int\!\! \mathrm{d}z\frac{T^z~
\Gamma 
\left[
	-z,1+z+\delta_2,1+z+\delta_4,
	-1-z-\delta_{0124},
	-1-z-\delta_{0234},
	2+z+\delta_{01234}
\right]
}{S^{2+z+\delta_{01234}}
\Gamma [-\delta_{001234},1+\delta_1,1+\delta_2,1+\delta_3,1+\delta_4]
} 
\,.
\label{one-r1-3}
\end{align}
The integrand of the higher order corrections 
has additional factors of $\alpha_i$
which can be expressed by
some shifts of $\delta_j\to\delta_j+1$.
Indeed, Eq.~\eqref{one-r1-1} can be expressed as
\begin{align}
I^{(1)}=\mathcal{T}^{(1)}_{\delta_1,\delta_2,\delta_3,\delta_4,\epsilon}
-\chi m_t^2
&\left(
\mathcal{P}_{1+\delta_1}^1\mathcal{T}^{(1)}_{1+\delta_1,\delta_2,\delta_3,\delta_4,\epsilon}
+\mathcal{P}_{1+\delta_2}^1\mathcal{T}^{(1)}_{\delta_1,1+\delta_2,\delta_3,\delta_4,\epsilon}
\right.
\nonumber\\
&\left.
+\mathcal{P}_{1+\delta_3}^1\mathcal{T}^{(1)}_{\delta_1,\delta_2,1+\delta_3,\delta_4,\epsilon}
+\mathcal{P}_{1+\delta_4}^1\mathcal{T}^{(1)}_{\delta_1,\delta_2,\delta_3,1+\delta_4,\epsilon}
\right)
+\mathcal{O}(\chi ^2)
\,,
\label{one-r1-4}
\end{align}
where $\mathcal{P}_x^n=\Gamma(x+n)/\Gamma(x)$
is the Pochhammer symbol.
We call $\mathcal{T}^{(1)}_{\delta_1,\delta_2,\delta_3,\delta_4,\epsilon}$
a ``template integral"
since all the higher order terms
can be expressed
in terms of $\mathcal{T}^{(1)}_{\delta_1,\delta_2,\delta_3,\delta_4,\epsilon}$
by shifting $\delta_j$.

Since there is no singularity in $\delta_j$,
we can safely set $\delta_j=0$ in Eq.~\eqref{one-r1-4}.
Then, the leading order term is expressed as
\begin{align}
\mathcal{T}^{(1)}_{0,0,0,0,\epsilon}
&=
\left.
\int _{-1/2-i\infty}^{-1/2+i\infty} 
\frac{\mathrm{d}z~T^z}{S^{2+\epsilon +z}}
\frac{\Gamma [
	-z,1+z,1+z,
	-1-\epsilon-z,
	-1-\epsilon-z,
	2+\epsilon+z]}
{\Gamma (-2\epsilon )}
\right|_{\epsilon \simeq -1}
\,,
\label{one-r1-5}
\end{align}
where the technique explained in Appendix~\ref{ss:introMB} is used
to set the integration contour to a straight line.
The expression at $\epsilon \to 0$ is obtained
by using the package \texttt{MB.m}~\cite{Czakon:2005rk},
and the result is
\begin{align}
\mathcal{T}^{(1)}_{0,0,0,0,\epsilon}
&=
e^{i\pi\epsilon}
\frac{e^{-\epsilon \gamma_E}}{s^{2+\epsilon}v}
\left[
-\frac{4}{\epsilon^2}
+\frac{2h_0+2i\pi}{\epsilon}
+\frac{4\pi^2}{3}
+\mathcal{O}(\epsilon)
\right]
\,.
\label{one-r1-7}
\end{align}
The higher order terms 
are given by $\mathcal{T}^{(1)}_{1,0,0,0,\epsilon}$,
$\mathcal{T}^{(1)}_{0,1,0,0,\epsilon}$ etc,
and can be calculated in a similar way.

The other procedure to calculate the contribution of the all-hard region
is the following.
We return to the momentum representation~\eqref{one1}
and expand each propagator in $m_t$ as
\begin{align}
\frac{1}{m_t^2-\ell ^2}\to \sum_{n=0}^\infty \frac{(-m_t^2)^n}{(-\ell^2)^{n+1}}
\,.
\label{one-r1-8}
\end{align}
Then, the contribution can be expressed
in terms of the integral family
\begin{align}
J_{a_1,a_2,a_3,a_4}^\mathrm{massless}=
  \int\!\!
  \frac{d^d\ell}{i \pi^{d/2}}
  \frac{1}
  {(-\ell^2)^{a_1}
  \left( -(\ell+q_1)^2 \right)^{a_2}
    \left( -(\ell+q_1+q_2)^2 \right)^{a_3}
  \left( -(\ell-q_3)^2 \right)^{a_4}}
  \,
  \label{one-r1-9}
\end{align}
as
\begin{align}
I^{(1)}=
J_{1,1,1,1}^\mathrm{massless}
-m_t^2(J_{2,1,1,1}^\mathrm{massless}+J_{1,2,1,1}^\mathrm{massless}
+J_{1,1,2,1}^\mathrm{massless}+J_{1,1,1,2}^\mathrm{massless})
+\mathcal{O}(m_t^4)
\,.
\label{one-r1-10}
\end{align}
This corresponds to Eq.~\eqref{one-r1-4}.
Applying the IBP-reduction,
all the integrals appearing Eq.~\eqref{one-r1-10}
including higher order terms 
can be expressed by three master integrals
\begin{align}
J_{1,0,1,0}^\mathrm{massless},~
J_{0,1,0,1}^\mathrm{massless},~
J_{1,1,1,1}^\mathrm{massless}
\,.
\end{align}
The IBP-reduction of massless integrals
is computationally easy even at the two-loop level,
and thus very useful.
In the calculation of two-loop integrals,
we adopt this approach to calculate the contribution from the all-hard region.

%%%%%%%%%%%%%%%%%%%%%%%%%%%%%%%%%%%%%%%%%%%%%%%%%%%%%%%
\subsubsection*{Regions 2, 3, 4, 5}

The contribution of Region~2 is obtained 
by applying the second scaling of Eq.~\eqref{one-scale} and expanding in 
$\chi m_t^2,\chi\alpha_3,\chi\alpha_4,$
\begin{align}
I^{(2)}&=\int\! \mathfrak{D}^4\alpha~
  {\mathcal{U}^{(2)}}^{-d/2}
  e^{ -\mathcal{F}^{(2)}/\mathcal{U}^{(2)} }
	\left[
	1-\chi 
	\left(m_t^2\alpha_{34}
	+\frac{d}{2}\frac{\alpha_{34}}{\mathcal{U}^{(2)}}
	+S\frac{\alpha_1\alpha_3\alpha_{34}}{(\mathcal{U}^{(2)})^2}
	+T\frac{\alpha_2\alpha_4\alpha_{34}}{(\mathcal{U}^{(2)})^2}
	\right)
	\right]
	+\mathcal{O}(\chi^2)
\label{one-r2-1}
\,,
\end{align}
where $\mathcal{U}^{(2)}=\alpha_{12}$ and 
$\mathcal{F}^{(2)}=m_t^2\alpha_{12}\mathcal{U}^{(2)}+S\alpha_1\alpha_3+T\alpha_2\alpha_4$.
The integration over $\alpha_1,\dots,\alpha_4$ can be performed
using the relation~\eqref{form1} and a variant of it,
and the template integral of this region is
\begin{align}
\mathcal{T}^{(2)}_{\delta_1,\delta_2,\delta_3,\delta_4,\epsilon}
=
\frac{(m_t^2)^{-\epsilon-\delta_1-\delta_2}}
{S^{1+\delta_3}T^{1+\delta_4}}
\frac{\Gamma [\delta_1-\delta_3,\delta_2-\delta_4,\delta_1+\delta_2+\epsilon]}
{\Gamma [\delta_{12}-\delta_{34},1+\delta_1,1+\delta_2]}
\,.
\label{one-r2-2}
\end{align}
The higher order terms in Eq.~\eqref{one-r2-1}
which contain the inverse of $\mathcal{U}^{(2)}$
can be expressed by the shift $\epsilon\to\epsilon -1$
in the template integral.
Thus Eq.~\eqref{one-r2-1} is written as
\begin{align}
I^{(2)}&=
\mathcal{T}^{(2)}_{\delta_1,\delta_2,\delta_3,\delta_4,\epsilon}
+\chi m_t^2
(\mathcal{P}_{1+\delta_3}^1\mathcal{T}^{(2)}_{\delta_1,\delta_2,\delta_3+1,\delta_4,\epsilon}
+\mathcal{P}_{1+\delta_4}^1\mathcal{T}^{(2)}_{\delta_1,\delta_2,\delta_3,\delta_4+1,\epsilon})
\nonumber\\
&+\chi \frac{d}{2}
(\mathcal{P}_{1+\delta_3}^1\mathcal{T}^{(2)}_{\delta_1,\delta_2,\delta_3+1,\delta_4,\epsilon-1}
+\mathcal{P}_{1+\delta_4}^1\mathcal{T}^{(2)}_{\delta_1,\delta_2,\delta_3,\delta_4+1,\epsilon-1})
\nonumber\\
&+\chi S
(\mathcal{P}_{1+\delta_1}^1\mathcal{P}_{1+\delta_3}^2
\mathcal{T}^{(2)}_{\delta_1+1,\delta_2,\delta_3+2,\delta_4,\epsilon-2}
+\mathcal{P}_{1+\delta_1}^1\mathcal{P}_{1+\delta_3}^1\mathcal{P}_{1+\delta_4}^1
\mathcal{T}^{(2)}_{\delta_1+1,\delta_2,\delta_3+1,\delta_4+1,\epsilon-2})
\nonumber\\
&+\chi T
(\mathcal{P}_{1+\delta_2}^1\mathcal{P}_{1+\delta_3}^1\mathcal{P}_{1+\delta_4}^1
\mathcal{T}^{(2)}_{\delta_1,\delta_2+1,\delta_3+1,\delta_4+1,\epsilon-2}
+\mathcal{P}_{1+\delta_2}^1\mathcal{P}_{1+\delta_4}^2
\mathcal{T}^{(2)}_{\delta_1,\delta_2+1,\delta_3,\delta_4+2,\epsilon-2})
+\mathcal{O}(\chi^2)
\label{one-r2-3}
\,.
\end{align}
Recall that $\mathcal{P}_x^n=\Gamma(x+n)/\Gamma(x)$
is the Pochhammer symbol.

As mentioned in Subsection~\ref{ss:conv},
the result of the limits $\delta_j\to0$
depends on the order in which we take them.
For example,
when we take the sequence of limits
with ascending values of $j$,
we obtain
\begin{align}
\lim_{\epsilon,\delta_4,\delta_3,\delta_2,\delta_1\to0}
\mathcal{T}^{(2)}_{\delta_1,\delta_2,\delta_3,\delta_4,\epsilon}
=
\frac{e^{-\epsilon \gamma_E}}{s^{2}v(m_t^2)^{\epsilon}}
\frac{1}{\epsilon}
\left( \frac{1}{\delta_3}+\frac{1}{\delta_4}
-\log s-h_0+i\pi
\right)
+\mathcal{O}(\epsilon)
\label{one-r2-4}
\,,
\end{align}
whereas with descending values of $j$, we instead obtain
\begin{align}
\lim_{\epsilon,\delta_1,\delta_2,\delta_3,\delta_4\to0}
\mathcal{T}^{(2)}_{\delta_1,\delta_2,\delta_3,\delta_4,\epsilon}
=
\frac{e^{-\epsilon \gamma_E}}{s^{2}v(m_t^2)^{\epsilon}}
\left[
\frac{2}{\epsilon^2}
-\frac{1}{\epsilon}\left( \frac{1}{\delta_1}+\frac{1}{\delta_2} -2\log (m_t^2) \right)
-\frac{\pi^2}{6}
\right]
+\mathcal{O}(\epsilon)
\label{one-r2-5}
\,.
\end{align}
The order dependence is not problem, provided 
we use the same order throughout the calculation.
The artifacts caused by the $\delta_j$ will cancel
after we sum the contributions from all the relevant regions.

Due to the symmetries of the diagrams,
the template integrals of the other regions can be expressed
in terms of $\mathcal{T}^{(2)}_{\delta_1,\delta_2,\delta_3,\delta_4,\epsilon}$
with exchanged $\delta_j$ as
\begin{align}
\mathcal{T}^{(3)}_{\delta_1,\delta_2,\delta_3,\delta_4,\epsilon}
&= \mathcal{T}^{(2)}_{\delta_1,\delta_4,\delta_3,\delta_2,\epsilon}
\label{one-r3-1}
\\
\mathcal{T}^{(4)}_{\delta_1,\delta_2,\delta_3,\delta_4,\epsilon}
&= \mathcal{T}^{(2)}_{\delta_3,\delta_2,\delta_1,\delta_4,\epsilon}
\label{one-r4-1}
\\
\mathcal{T}^{(5)}_{\delta_1,\delta_2,\delta_3,\delta_4,\epsilon}
&= \mathcal{T}^{(2)}_{\delta_3,\delta_4,\delta_1,\delta_2,\epsilon}
\,,
\label{one-r5-1}
\end{align}
and when we take the ascending order of limits, we obtain
\begin{align}
\lim_{\epsilon,\delta_4,\delta_3,\delta_2,\delta_1\to0}
\mathcal{T}^{(3)}_{\delta_1,\delta_2,\delta_3,\delta_4,\epsilon}
&=
\frac{e^{-\epsilon \gamma_E}}{s^{2}v(m_t^2)^{\epsilon}}
\left[
\frac{1}{\epsilon^2}
+\frac{1}{\epsilon}\left( \log(m_t^2)-\log s+i\pi +\frac{1}{\delta_3}-\frac{1}{\delta_4} \right)
-\frac{\pi^2}{12}
\right]
+\mathcal{O}(\epsilon)
\label{one-r3-2}
\\
\lim_{\epsilon,\delta_4,\delta_3,\delta_2,\delta_1\to0}
\mathcal{T}^{(4)}_{\delta_1,\delta_2,\delta_3,\delta_4,\epsilon}
&=
\frac{e^{-\epsilon \gamma_E}}{s^{2}v(m_t^2)^{\epsilon}}
\left[
\frac{1}{\epsilon^2}
+\frac{1}{\epsilon}\left( \log(m_t^2)-h_0 -\frac{1}{\delta_3}+\frac{1}{\delta_4} \right)
-\frac{\pi^2}{12}
\right]
+\mathcal{O}(\epsilon)
\label{one-r4-2}
\\
\lim_{\epsilon,\delta_4,\delta_3,\delta_2,\delta_1\to0}
\mathcal{T}^{(5)}_{\delta_1,\delta_2,\delta_3,\delta_4,\epsilon}
&=
\frac{e^{-\epsilon \gamma_E}}{s^{2}v(m_t^2)^{\epsilon}}
\left[
\frac{2}{\epsilon^2}
+\frac{1}{\epsilon}\left( 2\log(m_t^2)-\frac{1}{\delta_3}-\frac{1}{\delta_4} \right)
-\frac{\pi^2}{6}
\right]
+\mathcal{O}(\epsilon)
\label{one-r5-2}
\,.
\end{align}

%%%%%%%%%%%%%%%%%%%%%%%%%%%%%%%%%%%%%%%%%%%%%%%%%%%%%%%
\subsubsection*{Sum of all Regions}

Summing Eqs.~\eqref{one-r1-7}, \eqref{one-r2-4}, \eqref{one-r3-2}, \eqref{one-r4-2}, \eqref{one-r5-2},
we obtain the leading term of Eq.~\eqref{one1}:
\begin{align}
\mathrm{Eq.~\eqref{one1}}=
e^{i\pi\epsilon}
\frac{e^{-\epsilon \gamma_E}}{s^{2+\epsilon}v}
\left\{
\pi^2-2
\left[ \log \left( \frac{s}{m_t^2} \right) -i\pi \right]
\left[ \log \left( \frac{s}{m_t^2} \right) +h_0 \right]
\right\}
+\mathcal{O}(m_H^2,m_t^2,\epsilon)
\,.
\end{align}
As mentioned, the result is $\delta_j$-independent.
There are 24 possible ways to order $\delta_1,\delta_2,\delta_3,\delta_4$
in taking the limit,
and we have confirmed that the result is the same for all of the orderings.
Since the original integral is finite in the limit $\epsilon\to0$,
the poles of $\epsilon$ in the individual contributions from each region
cancel.

%%%%%%%%%%%%%%%%%%%%%%%%%%%%%%%%%%%%%%%%%%%%%%%%%%%%%%%
\subsection{Higher Order Terms in $m_t$}
\label{ss:one-higher}

The method of regions can be used to obtain
a series expansion up to arbitrary order in the soft parameters, $m_H$ or $m_t$.
For example, Eqs.~\eqref{one-r1-4},~\eqref{one-r2-3}
includes the contributions up to $\mathcal{O}(\chi )$.
In general, the higher order terms can be expressed in terms of
the template integrals.
Therefore, in principle,
it is straightforward to calculate higher order terms up to arbitrary order, 
once the template integrals have been obtained.

However, the number of integrals to calculate increases rapidly
with the order of $\chi$,
and this makes it hard to calculate higher order terms.
Especially in the two-loop case,
some of the template integrals contain 
multi-dimensional Mellin-Barnes integrals,
which are not so easy to solve.
Furthermore, the number of integrals 
increases more rapidly than in the case of the one-loop calculation.
Therefore it is better 
to use another method to calculate the higher-order corrections.

The use of differential equations solves this issue~
\cite{Melnikov:2016qoc,Kudashkin:2017skd,Davies:2018ood,Davies:2018qvx}.
We use the differential equation with respect to $m_t^2$
to obtain higher order corrections in $m_t^2$.
Since we know that the integral has the form
\begin{align}
\mathrm{Eq.~\eqref{one-alpha2}}=
\sum_{n_1} \sum_{n_2}
c_{n_1,n_2}(S,T)
(m_t^2)^{n_1}
\left( \log m_t^2 \right)^{n_2}
\,,
\label{diffeq1}
\end{align}
the set of differential equations reduces
to a set of linear relations of $c_{n_1,n_2}(S,T)$
which simplifies the problem a lot.
In this sense,
the leading order terms which we calculate in the previous subsections 
play the role of the boundary conditions of the differential equations.

%%%%%%%%%%%%%%%%%%%%%%%%%%%%%%%%%%%%%%%%%%%%%%%%%%%%%%%
\subsection{Integrals with Fewer Lines}
\label{ss:fewer}

Once we have calculated the box integral,
there are several shortcuts to
calculate integrals with fewer lines
such as the triangle integral and the self energy integral.
Let us consider the $s$-channel triangle diagram,
$J_{1,1,1,0}$, as an example.

The alpha representation of $J_{1,1,1,0}$
is obtained by
setting $\alpha_4\to 0$ in Eq.~\eqref{one-uf},
since the forth propagator is absent in $J_{1,1,1,0}$.
If we use \texttt{asy2.1.m} to reveal the relevant scalings
for $J_{1,1,1,0}$, we obtain
\begin{align}
(0,0,0),~(0,0,1),~(1,0,0),
\end{align}
however, we do not have to do that.
We do not have to derive the template integrals for $J_{1,1,1,0}$
because they are derived from the template integrals of $J_{1,1,1,1}$.

Using the fact that 
the $\delta_j$-dependence 
of the alpha representation is expressed
by the replacement of $a_i\to 1+\delta_j$,
the triangle integral $J_{1,1,1,0}$
can be expressed by the limit $\delta_4\to-1$.
Therefore by taking the limit $\delta_4\to-1$
to the template integral of $J_{1,1,1,1}$,
one can obtain the template integral for $J_{1,1,1,0}$.
For example,
\begin{align}
\lim_{\delta_4\to-1}
\mathrm{Eq.~\eqref{one-r2-2}}
=
\frac{(m_t^2)^{-\epsilon-\delta_1-\delta_2}}
{S^{1+\delta_3}}
\frac{\Gamma [\delta_1-\delta_3,1+\delta_2,\delta_1+\delta_2+\epsilon]}
{\Gamma [1+\delta_{12}-\delta_{3},1+\delta_1,1+\delta_2]}
\,,
\label{one-few1}
\end{align}
and this is the template integral for the region $(0,0,1)$.
Sometimes the limit vanishes due to a suppression factor $1/\Gamma(1+\delta_4)$.
For example,
\begin{align}
\lim_{\delta_4\to-1}
\mathrm{Eq.~\eqref{one-r3-1}}
=
0
\,.
\label{one-few2}
\end{align}
This fact is reasonable because the number of relevant regions 
for $J_{1,1,1,0}$ is smaller than for $J_{1,1,1,1}$.
Two of the four soft template integrals are non-vanishing after taking the limit,
and they are the two soft template integrals of $J_{1,1,1,0}$.

%%%%%%%%%%%%%%%%%%%%%%%%%%%%%%%%%%%%%%%%%%%%%%%%%%%%%%%
\section{Two-Loop Planar Diagrams}
\label{ss:twoPL}

\begin{figure}
\centering
%\begin{tabular}{cc}
%\begin{minipage}{0.45\hsize}
%\centering
%\begin{tikzpicture}[scale=0.5]
%\draw [ultra thick] (0,0)--(4,0)--(4,4)--(0,4)--(0,0);
%\draw [dotted, ultra thick] (0,0)--(-1,-1);
%\draw [dotted, ultra thick] (4,0)--(5,-1);
%\draw [dotted, ultra thick] (4,4)--(5,5);
%\draw [dotted, ultra thick] (0,4)--(-1,5);
%\draw [dotted, ultra thick] (2,0)--(2,4);
%\draw (1.1,4) node [above] {$p_1$};
%\draw (0,2) node [left] {$p_2$};
%\draw (1.1,0) node [below] {$p_3$};
%\draw (3.1,0) node [below] {$p_4$};
%\draw (4.1,2) node [right] {$p_5$};
%\draw (3.1,4) node [above] {$p_6$};
%\draw (2.1,2) node [right] {$p_7$};
%\end{tikzpicture}
%\end{minipage}
%\begin{minipage}{0.45\hsize}
%\centering
%\begin{tikzpicture}[scale=0.5]
%\draw [dotted, ultra thick] (2,0)--(0,0)--(0,4)--(2,4);
%\draw [ultra thick] (2,0)--(4,0)--(4,4)--(2,4)--(2,0);
%\draw [dotted, ultra thick] (0,0)--(-1,-1);
%\draw [dotted, ultra thick] (4,0)--(5,-1);
%\draw [dotted, ultra thick] (4,4)--(5,5);
%\draw [dotted, ultra thick] (0,4)--(-1,5);
%\draw (1.1,4) node [above] {$p_1$};
%\draw (0,2) node [left] {$p_2$};
%\draw (1.1,0) node [below] {$p_3$};
%\draw (3.1,0) node [below] {$p_4$};
%\draw (4.1,2) node [right] {$p_5$};
%\draw (3.1,4) node [above] {$p_6$};
%\draw (2.1,2) node [right] {$p_7$};
%\end{tikzpicture}
%\end{minipage}
%\end{tabular}
\includegraphics[width=0.8\textwidth]{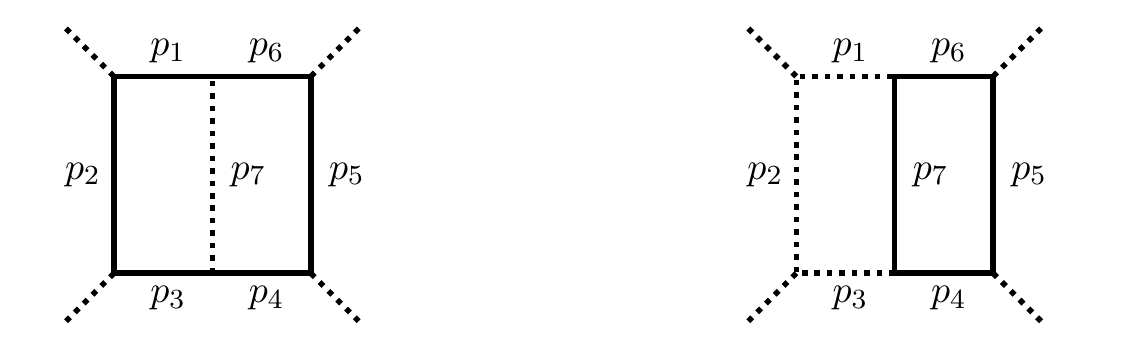}
\caption{
The two-loop massive planar diagrams
defined in Eq.~\eqref{pl-1} (left) and Eq.~\eqref{pl-2} (right).}
\label{fig:two-pl}
\end{figure}

We consider the following Feynman integral families
\begin{align}
J^\mathrm{PL1}_{a_1,a_2,a_3,a_4,a_5,a_6,a_7}&=
\int\!\!
\frac{\mathrm{d}^d\ell_1}{i \pi^{d/2}}
\frac{\mathrm{d}^d\ell_2}{i \pi^{d/2}}
\frac{1}{(-p_7^2)^{a_7}}
\prod _{n=1}^6 \frac{1}{(m_t^2-p_n^2)^{a_n}}
\label{pl-1}
\\
J^\mathrm{PL2}_{a_1,a_2,a_3,a_4,a_5,a_6,a_7}&=
\int\!\!
\frac{\mathrm{d}^d\ell_1}{i \pi^{d/2}}
\frac{\mathrm{d}^d\ell_2}{i \pi^{d/2}}
\prod _{n=1,2,3,7} \frac{1}{(m_t^2-p_n^2)^{a_n}}
\prod _{n=4}^6 \frac{1}{(-p_n^2)^{a_n}}
\label{pl-2}
\end{align}
where the momenta of the lines are 
\begin{align}
\{p_1,p_2,p_3,p_4,p_5,p_6,p_7\}
=\{\ell_1+q_{1},
\ell_1,
\ell_1-q_2,
\ell_2-q_2,
\ell_{2}+q_{13},
\ell_2+q_1,
\ell_1-\ell_2\} \,.
\label{pl-3}
\end{align}
Recall that $q_{13}=q_1+q_3$.
We consider the integrals 
$I^\mathrm{PL1}=J^\mathrm{PL1}_{1,1,1,1,1,1,1}$
and $I^\mathrm{PL2}=J^\mathrm{PL2}_{1,1,1,1,1,1,1}$
whose diagrammatic representations are shown in Fig.~\ref{fig:two-pl}.

The $m_H$-expansion can be performed in the same way as in Subsection~\ref{ss:one-higgs},
and the alpha representations of the integrals after the $m_H$-expansion are given by
\begin{align}
&\mathcal{U}^\mathrm{PL1}=
\mathcal{U}^\mathrm{PL2}=
\alpha_{123}\alpha_{456}+\alpha_{123456}\alpha_7\,,
\\
&\mathcal{F}^\mathrm{PL1}=m_t^2\alpha_{123456}\mathcal{U}^\mathrm{PL1}
+S\left[ \alpha_1\left( \alpha_4\alpha_{67}+\alpha_3\alpha_{4567} \right)
+\alpha_6\left( \alpha_{23}\alpha_4+\alpha_{34}\alpha_7\right) \right]
+T\alpha_2\alpha_5\alpha_7\,,
\\
&\mathcal{F}^\mathrm{PL2}=m_t^2\alpha_{1237}\mathcal{U}^\mathrm{PL2}
+S\left[ \alpha_1\left( \alpha_4\alpha_{67}+\alpha_3\alpha_{4567} \right)
+\alpha_6\left( \alpha_{23}\alpha_4+\alpha_{34}\alpha_7\right) \right]
+T\alpha_2\alpha_5\alpha_7
\,.
\end{align}
Conceptually there is no difference
between 
the procedure of applying the method of regions
to these integrals
and the example discussed 
in Section~\ref{ss:one}.
In particular,
the property of the $\mathcal{F}$-function that
it is positive definite in the $u$-channel is the same
[cf. the text below Eq.~\eqref{one-uf2}].
The only new ingredient is that
now the template integrals are expressed by
at most two-dimensional Mellin-Barnes integrals,
which are not trivial to solve.
However, their calculation is a subset of
the calculation of the non-planar integrals,
so we do not describe it here
[cf. Subsection~\ref{ss:mb}].
Therefore we briefly summarize the important ingredients
of the two-loop planar integrals
in this section.

\subsection*{Double Massive Box Diagram}
By using the package \texttt{asy2.1.m},
we reveal thirteen relevant scalings:
\begin{align}
&(0,0,0,0,0,0,0)
,
(0,0,1,0,0,1,1)
,
(0,0,1,1,0,0,1)
,
(0,0,1,1,1,0,0)
,
(0,1,1,1,0,0,0)
,\nonumber\\
&(1,0,0,0,0,1,1)
,
(1,0,0,0,1,1,0)
,
(1,0,0,1,0,0,1)
,
(1,1,0,0,0,1,0)
,
(0,0,1,1,1,1,1)
,\nonumber\\
&(1,0,0,1,1,1,1)
,
(1,1,1,0,0,1,1)
,
(1,1,1,1,0,0,1)
\,.
\end{align}
The template integrals of these regions are
summarized in Appendix~\ref{app:temp},
and can be found in the ancillary file to this paper~\cite{anci}.
The result of this integral at
the leading order in $m_t$, up to $\mathcal{O}(\epsilon )$, is
given by
\begin{align}
I^\mathrm{PL1}=\sum_{n_1=0}^1\sum_{n_2=0}^{n_1+4}
\frac{d_{n_1,n_2}}{v} \epsilon ^{n_1} \log ^{n_2} (m_t)
+\mathcal{O}(m_t,\epsilon ^2)
\,,
\end{align}
where 
the coefficients $d_{n_1,n_2}$ are given by
\begin{align}
d_{0,4}&=16\,,\qquad
d_{0,3}=-\frac{64 h_0}{3}+\frac{32 i \pi }{3}\,,\qquad
d_{0,2}=8 h_0^2-16 i \pi  h_0-\frac{8 \pi ^2}{3}\,,\nonumber\\
d_{0,1}&=8 i \pi  h_0^2+\frac{16 \pi ^2 h_0}{3}-8 \zeta_{3}+\frac{8 i \pi ^3}{3}\,,\nonumber\\
d_{0,0}&=16 h_0 h_3-\frac{h_0^4}{3}-\frac{4}{3} i \pi  h_0^3-4 h_0^2 h_2-2 \pi ^2 h_0^2-8 i \pi  h_0 h_2+4 h_0 \zeta_{3}-\frac{4}{3} i \pi ^3 h_0+16 i \pi  h_3-24 h_4-\frac{7 \pi ^4}{15}\,,\nonumber\\
d_{1,5}&=-\frac{128}{3}\,,\qquad
d_{1,4}=\frac{152 h_0}{3}-56 i \pi\,,\qquad
d_{1,3}=-16 h_0^2+\frac{208 i \pi  h_0}{3}+\frac{160 \pi ^2}{9}\,,\nonumber\\
d_{1,2}&=\frac{4 h_0^3}{3}+4 h_0^2 h_1-20 i \pi  h_0^2+8 i \pi  h_0 h_1-8 h_0 h_2-28 \pi ^2 h_0-8 i \pi  h_2+8 h_3-16 \zeta_{3}-\frac{4 i \pi ^3}{3}\,,\nonumber\\
d_{1,1}&=-\frac{4 h_0^4}{3}-\frac{8 h_0^3 h_1}{3}-4 i \pi  h_0^3-4 i \pi  h_0^2 h_1+\frac{28 \pi ^2 h_0^2}{3}-8 i \pi  h_0 h_2+16 h_0 h_3+24 i \pi  h_3-32 h_4\nonumber\\
&-\frac{8}{3} \pi ^2 h_0 h_1+16 h_0 \zeta_{3}-\frac{28}{3} i \pi ^3 h_0+\frac{8 \pi ^2 h_2}{3}+\frac{218 \pi ^4}{45}\,,\nonumber\\
d_{1,0}&=\frac{h_0^5}{2}+\frac{h_0^4 h_1}{2}+6 h_0^3 h_2+6 h_0^2 h_1 h_2+4 h_0^2 h_{21}-20 h_0^2 h_3-4 h_0 h_2^2-12 h_0 h_{22}\nonumber\\
&+\frac{11}{6} i \pi  h_0^4+\frac{2}{3} i \pi  h_0^3 h_1+18 i \pi  h_0^2 h_2+12 i \pi  h_0 h_1 h_2-24 h_0 h_1 h_3+36 h_1 h_4-4 i \pi  h_2^2+4 h_2 h_3\nonumber\\
&+\frac{\pi ^2 h_0^3}{9}+\frac{5}{3} \pi ^2 h_0^2 h_1-6 \pi ^2 h_0 h_2+8 i \pi  h_0 h_{21}-32 i \pi  h_0 h_3-24 i \pi  h_1 h_3-12 i \pi  h_{22}-16 i \pi  h_4\nonumber\\
&-4 h_0^2 \zeta_{3}+5 i \pi ^3 h_0^2+2 i \pi ^3 h_0 h_1-\frac{157 \pi ^4 h_0}{90}-\frac{2 \pi ^4 h_1}{5}+\frac{10}{3} i \pi ^3 h_2+\frac{26 \pi ^2 h_3}{3}+\frac{61 i \pi ^5}{90}\nonumber\\
&+24 h_0 h_1 \zeta_{3}+24 i \pi  h_1 \zeta_{3}-56 h_2 \zeta_{3}+4 (4 h_{23}+7 h_{32}+21 h_5-13 \zeta_{5})+6 \pi ^2 \zeta_{3}\,.
\end{align}

\subsection*{Single Massive Double Box Diagram}
By using the package \texttt{asy2.1.m},
we reveal ten relevant scalings:
\begin{align}
&(0,0,0,0,0,0,0)
,
(0,0,1,1,1,0,0)
,
(0,1,1,1,0,0,0)
,
(1,0,0,0,1,1,0)
,
\nonumber\\
&(1,1,0,0,0,1,0)
,
(0,0,1,1,1,1,1)
,
(1,0,0,1,1,1,1)
,
(1,1,1,1,1,1,0)
\,.
\end{align}
The template integrals of these regions are
summarized in Appendix~\ref{app:temp},
and can be found in the ancillary file to this paper~\cite{anci}.
The result of this integral at
the leading order in $m_t$, up to $\mathcal{O}(\epsilon )$, is
\begin{align}
I^\mathrm{PL2}=\sum_{n_1=-2}^1\sum_{n_2=0}^{n_1+4}
\frac{d_{n_1,n_2}}{v} \epsilon ^{n_1} \log ^{n_2} (m_t)
+\mathcal{O}(m_t,\epsilon ^2)
\,,
\end{align}
where the coefficients $d_{n_1,n_2}$ are now given by
\begin{align}
d_{-2,2}&=8\,,\qquad
d_{-2,1}=-4 h_0+4 i \pi\,,\qquad
d_{-2,0}=-\pi ^2-2 i \pi  h_0\,,\nonumber\\
d_{-1,3}&=-\frac{32}{3}\,,\qquad
d_{-1,2}=-4 h_0-20 i \pi\,,\qquad
d_{-1,1}=4 h_0^2+4 i \pi  h_0+\frac{20 \pi ^2}{3}\,,\nonumber\\
d_{-1,0}&=\frac{h_0^3}{3}+h_0^2 h_1+3 i \pi  h_0^2+2 i \pi  h_0 h_1-2 h_0 h_2-\frac{4 \pi ^2 h_0}{3}-2 i \pi  h_2+2 h_3-14 \zeta_{3}+2 i \pi ^3\,,\nonumber\\
d_{0,4}&=8\,,\qquad
d_{0,3}=\frac{32 h_0}{3}+\frac{80 i \pi }{3}\,,\qquad
d_{0,2}=-4 h_0^2+8 i \pi  h_0-\frac{70 \pi ^2}{3}\,,\nonumber\\
d_{0,1}&=-4 i \pi  h_0^2+4 \pi ^2 h_0+20 \zeta_{3}-\frac{10 i \pi ^3}{3}\,,\nonumber\\
d_{0,0}&=-\frac{5 h_0^4}{6}-\frac{4 h_0^3 h_1}{3}-\frac{h_0^2 h_1^2}{2}-3 h_0^2 h_2+2 h_0 h_1 h_2-2 h_0 h_{21}+20 h_0 h_3+\frac{h_2^2}{2}-h_{22}\nonumber\\
&-\frac{10}{3} i \pi  h_0^3-4 i \pi  h_0^2 h_1-i \pi  h_0 h_1^2-6 i \pi  h_0 h_2+2 i \pi  h_1 h_2-2 h_1 h_3-2 i \pi  h_{21}+20 i \pi  h_3-34 h_4\nonumber\\
&+\frac{11 \pi ^2 h_0^2}{6}+\frac{7}{3} \pi ^2 h_0 h_1+14 h_0 \zeta_{3}-\frac{7}{3} i \pi ^3 h_0+2 h_1 \zeta_{3}-\frac{1}{3} i \pi ^3 h_1-\frac{7 \pi ^2 h_2}{3}+24 i \pi  \zeta_{3}+\frac{259 \pi ^4}{180}\,,\nonumber\\
d_{1,5}&=-\frac{64}{15}\,,\qquad
d_{1,4}=-12 h_0-\frac{68 i \pi }{3}\,,\qquad
d_{1,3}=\frac{8 h_0^2}{3}-\frac{56 i \pi  h_0}{3}+\frac{272 \pi ^2}{9}\,,\nonumber\\
d_{1,2}&=\frac{2 h_0^3}{3}+2 h_0^2 h_1+6 i \pi  h_0^2+4 i \pi  h_0 h_1-4 h_0 h_2+\frac{16 \pi ^2 h_0}{3}-4 i \pi  h_2+4 h_3-\frac{220 \zeta_{3}}{3}+\frac{40 i \pi ^3}{3}\,,\nonumber\\
d_{1,1}&=-\frac{2 h_0^4}{3}-\frac{4 h_0^3 h_1}{3}-2 i \pi  h_0^3-2 i \pi  h_0^2 h_1-\frac{8 \pi ^2 h_0^2}{3}-4 i \pi  h_0 h_2+8 h_0 h_3+12 i \pi  h_3-16 h_4\nonumber\\
&-\frac{4}{3} \pi ^2 h_0 h_1+\frac{44 h_0 \zeta_{3}}{3}-\frac{16}{3} i \pi ^3 h_0+\frac{4 \pi ^2 h_2}{3}-\frac{176 i \pi  \zeta_{3}}{3}+\frac{5 \pi ^4}{18}\,,\nonumber\\
d_{1,0}&=\frac{43 h_0^5}{60}+\frac{5 h_0^4 h_1}{4}+\frac{2 h_0^3 h_1^2}{3}+\frac{17 h_0^3 h_2}{3}+\frac{h_0^2 h_1^3}{6}+5 h_0^2 h_1 h_2-h_0 h_1^2 h_2\nonumber\\
&+5 h_0^2 h_{21}+2 h_0 h_1 h_{21}-3 h_0 h_2^2-\frac{h_1 h_2^2}{2}+h_1 h_{22}+\frac{h_2 h_{21}}{3}-\frac{h_{212}}{3}\nonumber\\
&-22 h_0^2 h_3-28 h_0 h_1 h_3-2 h_0 (h_{211}+7 h_{22})+h_1^2 h_3+\frac{7 h_2 h_3}{3}-h_{221}+\frac{53 h_{23}}{3}\nonumber\\
&+\frac{13}{4} i \pi  h_0^4+\frac{13}{3} i \pi  h_0^3 h_1+2 i \pi  h_0^2 h_1^2+\frac{1}{3} i \pi  h_0 h_1^3+46 h_1 h_4+33 h_{32}+98 h_5\nonumber\\
&+17 i \pi  h_0^2 h_2+10 i \pi  h_0 h_1 h_2+10 i \pi  h_0 h_{21}-i \pi  h_1^2 h_2+2 i \pi  h_1 h_{21}-3 i \pi  h_2^2-2 i \pi  (h_{211}+7 h_{22})\nonumber\\
&-\frac{22}{9} \pi ^2 h_0^3-3 \pi ^2 h_0^2 h_1-\frac{7}{6} \pi ^2 h_0 h_1^2-5 \pi ^2 h_0 h_2-40 i \pi  h_0 h_3-28 i \pi  h_1 h_3-8 i \pi  h_4\nonumber\\
&+\frac{10}{3} i \pi ^3 h_0^2+\frac{4}{3} i \pi ^3 h_0 h_1+\frac{1}{6} i \pi ^3 h_1^2+\frac{7}{3} \pi ^2 h_1 h_2+5 i \pi ^3 h_2-\frac{7 \pi ^2 h_{21}}{3}+16 \pi ^2 h_3\nonumber\\
&-9 h_0^2 \zeta_{3}+26 h_0 h_1 \zeta_{3}-\frac{599 \pi ^4 h_0}{180}-h_1^2 \zeta_{3}-\frac{161 \pi ^4 h_1}{180}-64 h_2 \zeta_{3}-\frac{47 i \pi ^5}{90}\nonumber\\
&-\frac{32}{3} i \pi  h_0 \zeta_{3}+26 i \pi  h_1 \zeta_{3}+29 \pi ^2 \zeta_{3}-121 \zeta_{5}\,.
\end{align}

%%%%%%%%%%%%%%%%%%%%%%%%%%%%%%%%%%%%%%%%%%%%%%%%%%%%%%%
\section{Two-Loop Non-Planar Diagram}
\label{ss:two}

\begin{figure}
\centering
%\begin{tabular}{cc}
%\begin{minipage}{0.45\hsize}
%\centering
%%%%%%%%%%%%%%%%%%%
%\begin{tikzpicture}[scale=0.5]
%\draw[ultra thick] (0,4)--(4,4)--(4,0)--(2,0)--(0,4);
%\draw[ultra thick, dotted] (2,0)--(0,0)--(2,4);
%\draw [dotted, ultra thick] (0,0)--(-1,-1);
%\draw [dotted, ultra thick] (4,0)--(5,-1);
%\draw [dotted, ultra thick] (4,4)--(5,5);
%\draw [dotted, ultra thick] (0,4)--(-1,5);
%\draw (0.5,1) node [left] {$p_1$};
%\draw (1.1,0) node [below] {$p_2$};
%\draw (0.5,3) node [left] {$p_3$};
%\draw (1.1,4) node [above] {$p_4$};
%\draw (3.1,4) node [above] {$p_5$};
%\draw (4,2) node [right] {$p_6$};
%\draw (3.1,0) node [below] {$p_7$};
%\end{tikzpicture}
%%%%%%%%%%%%%%%%%%%
%\end{minipage}
%\begin{minipage}{0.45\hsize}
%\centering
%%%%%%%%%%%%%%%%%%%
%\begin{tikzpicture}[scale=0.5]
%\draw[ultra thick] (4-1.1,4-2.2)--(2,0)--(4,0)--(2,4)--(4,4)--(4-0.9,4-1.8);
%\draw[ultra thick, dotted] (2,0)--(0,0)--(0,4)--(2,4);
%\draw [dotted, ultra thick] (0,0)--(-1,-1);
%\draw [dotted, ultra thick] (4,0)--(5,-1);
%\draw [dotted, ultra thick] (4,4)--(5,5);
%\draw [dotted, ultra thick] (0,4)--(-1,5);
%\draw (3.5,1) node [right] {$p_4$};
%\draw (1.1,0) node [below] {$p_7$};
%\draw (3.5,3) node [right] {$p_2$};
%\draw (1.1,4) node [above] {$p_5$};
%\draw (3.1,4) node [above] {$p_1$};
%\draw (0,2) node [left] {$p_6$};
%\draw (3.1,0) node [below] {$p_3$};
%\end{tikzpicture}
%%%%%%%%%%%%%%%%%%%
%\end{minipage}
%\end{tabular}
\includegraphics[width=0.8\textwidth]{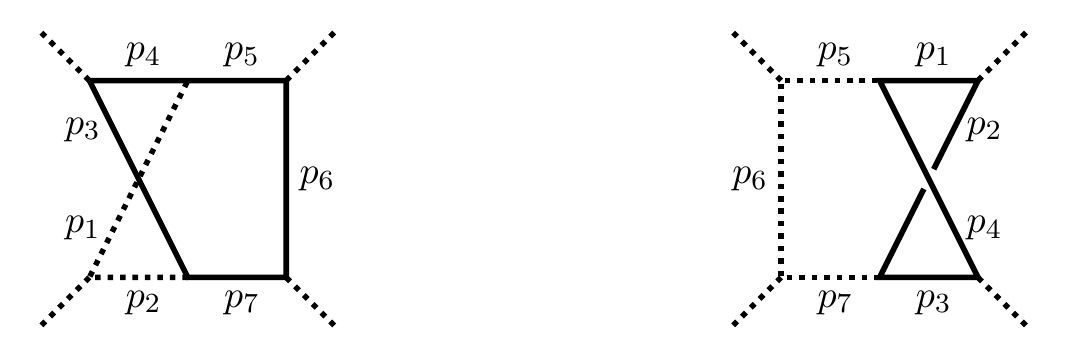}
\caption{
The two-loop massive non-planar diagrams
defined in Eq.~\eqref{two-1} (left) and Eq.~\eqref{two-npl2-1} (right).}
\label{fig:two}
\end{figure}

For the two-loop massive non-planar diagrams,
we consider the following two Feynman integral families  
\begin{align}
J^\mathrm{NPL1}_{a_1,a_2,a_3,a_4,a_5,a_6,a_7}&=
\int\!\!
\frac{\mathrm{d}^d\ell_1}{i \pi^{d/2}}
\frac{\mathrm{d}^d\ell_2}{i \pi^{d/2}}
\prod _{n=1}^2 \frac{1}{(-p_n^2)^{a_n}}
\prod _{n=3}^7 \frac{1}{(m_t^2-p_n^2)^{a_n}}
\label{two-1}
\end{align}
where the momenta of the lines are given by
\begin{align}
\{p_1,p_2,p_3,p_4,p_5,p_6,p_7\}
=\{\ell_1+q_{12},
\ell_1-q_3,
\ell_{12}+q_{2\bar3},
\ell_{12}+q_2,
\ell_2-q_1,
\ell_2,
\ell_2+q_2\} \,,
\label{two-2}
\end{align}
and 
\begin{align}
J^\mathrm{NPL2}_{a_1,a_2,a_3,a_4,a_5,a_6,a_7}&=
\int\!\!
\frac{\mathrm{d}^d\ell_1}{i \pi^{d/2}}
\frac{\mathrm{d}^d\ell_2}{i \pi^{d/2}}
\prod _{n=1}^4 \frac{1}{(m_t^2-p_n^2)^{a_n}}
\prod _{n=5}^7 \frac{1}{(-p_n^2)^{a_n}}
\label{two-npl2-1}
\end{align}
where the momenta of the lines are 
\begin{align}
\{p_1,p_2,p_3,p_4,p_5,p_6,p_7\}
=\{\ell_1,
\ell_1+q_3,
\ell_{12}+q_{23},
\ell_{12}-q_1
\ell_2-q_1,
\ell_2,
\ell_2+q_2\} \,.
\label{two-npl2-2}
\end{align}
Recall that $q_{2\bar 3}=q_2-q_3$ and $q_{23}=q_2+q_3$.
We consider 
$I^\mathrm{NPL1}=J^\mathrm{NPL1}_{1,1,1,1,1,1,1}$ as an example
in this section.
The template integrals of 
$I^\mathrm{NPL2}=J^\mathrm{NPL2}_{1,1,1,1,1,1,1}$
is provided in Appendix~\ref{app:temp}.
The Feynman diagrams corresponding
$I^\mathrm{NPL1}$
and 
$I^\mathrm{NPL2}$
are illustrated in Fig.~\ref{fig:two}.

In Subsection~\ref{ss:one-higgs} we showed that the expansion in $m_H$ 
can be obtained by
the naive Taylor expansion of the integrand.
This holds also in this case,
and the $m_H$-expansion is straightforward.
Therefore we again consider only integrals with massless external legs.
The alpha representation of
our non-planar integral is
\begin{align}
I^\mathrm{NPL1}&\myeq
\int\! \mathfrak{D}^7\alpha^\delta~
\mathcal{U}^{-d/2}
e^{ -\mathcal{F}/\mathcal{U} }
\,,
\label{two-4}
\end{align}
where 
\begin{align}
\mathcal{U}&=\alpha_{12}\alpha_{34567}+\alpha_{34}\alpha_{567}
\label{two-5}
\\
\mathcal{F}&=
m_t^2\alpha_{34567}\mathcal{U}
+S\left( \alpha_1\alpha_7\alpha_{45}+\alpha_2\alpha_5\alpha_{37}+\alpha_5\alpha_7\alpha_{34} \right)
+T\alpha_1\alpha_3\alpha_6
+U\alpha_2\alpha_4\alpha_6
\label{two-6}
\,.
\end{align}

%%%%%%%%%%%%%%%%%%%%%%%%%%%%%%%%%%%%%%%%%%%%%%%%%%%%%%%
\subsection{Relevant Scaling}
\label{ss:two-scale}

A new feature appears in Eq.~\eqref{two-6}:
if we impose the relation $S+T+U=0$,
we obtain 
\begin{align}
\mathcal{F}
\myeq
m_t^2\alpha_{34567}\mathcal{U}
+S\left( \alpha_1\alpha_7\alpha_{45}+\alpha_2\alpha_5\alpha_{37}+\alpha_5\alpha_7\alpha_{34}
-\alpha_2\alpha_4\alpha_6 \right)
+T(\alpha_1\alpha_3\alpha_6-\alpha_2\alpha_4\alpha_6)
\,,
\label{two-f2}
\end{align}
and the sign of $\mathcal{F}$ becomes indefinite.
In such a case,
it is not guaranteed that the method to reveal the relevant scalings
works properly~\cite{Jantzen:2012mw}.
An idea to solve this problem is 
to perform a proper change of variables
and decompose the integration domain
such that $\mathcal{F}$ is positive-definite~\cite{Jantzen:2012mw}.
However in our case,
this approach does not resolve the indefinite sign of $\mathcal{F}$
since there is no simple change of variables 
to make $\mathcal{F}$ positive-definite.

The solution to this problem is to keep $S, T, U$ as independent variables.
It is obvious that Eq.~\eqref{two-6} is positive definite in this case,
and we can apply the method of regions,
expand the integrand, and express the result in terms of Mellin-Barnes integrals
in terms of the positive Mandelstam variables.
The procedure to obtain expression~\eqref{two-6} is the following:
we first compute $\mathcal{F}$ respecting the original definition of 
the Mandelstam variables~\eqref{stu}.
At this point,
there are some redundant terms
in $\mathcal{F}$
such as $(S+T+U)\alpha_2\alpha_3\alpha_6$.
We minimize the number of terms,
under the condition that $\mathcal{F}$ remains positive definite.
The resulting $\mathcal{F}$ is unique.

There are two commands in the package \texttt{asy2.1.m}
to reveal the relevant regions.
The first is \texttt{AlphaRepExpand},
which accepts a set of propagators and replacement rules as input.
The other is \texttt{WilsonExpand},
which accepts the Symanzik polynomials as input.
Here we must use \texttt{WilsonExpand}
since the conventional routine used in \texttt{AlphaRepExpand}
either eliminates $U$ completely or keeps $U$ completely,
whereas 
we want to eliminate $U$ partially,
as explained above.
There is an option \texttt{Preresolve}
in \texttt{AlphaRepExpand}
which makes it attempt some changes of variables
to make $\mathcal{F}$ positive definite,
but in our cases this option did not solve the problem.

With this setup and the hierarchy~\eqref{sca2},
we obtain the following fourteen relevant scalings\footnote{
In fact, it turns out that 
the correct scalings~\eqref{two-scalings}
can be obtained 
by using Eq.~\eqref{two-f2}
or by using \texttt{AlphaRepExpand},
provided suitable values for $S$ and $T$ are chosen 
(e.g. $S=1, T=1$),
such that $U\not =0$.
This may be because 
there is no cancellation between two hard-scaling terms
resulting in a soft-scaling term in Eq.~\eqref{two-f2}.
However, this observation is made in hindsight,
since in principle it is not guaranteed that the regions are found correctly.
One could also use 
\texttt{ASPIRE}~\cite{Ananthanarayan:2018tog}
instead of \texttt{asy2.1.m}
and obtain the correct scalings,
if one similarly chooses suitable values for $S$ and $T$.
}
\begin{align}
&
(0,0,0,0,0,0,0),
(0,\frac{1}{2},\frac{1}{2},0,0,\frac{1}{2},1), 
(\frac{1}{2},0,0,\frac{1}{2},1,\frac{1}{2},0), 
(0,0,0,1,1,1,0)
\nonumber\\
&
(0,0,1,0,0,1,1),
(0,1,0,0,0,1,1),
(0,1,1,0,0,0,1),
(1,0,0,0,1,1,0),
(1,0,0,1,1,0,0),
\nonumber\\
&
(1,1,0,0,0,0,1),
(1,1,0,0,1,0,0),
(1,1,0,0,1,1,1),
(1,1,1,1,0,0,1),
(1,1,1,1,1,0,0)
\,.
\label{two-scalings}
\end{align}

In the case of the planar integrals~\cite{Davies:2018ood},
the scalings consist of only 0 and 1.
Here, we additionally have a scaling $(0, \frac{1}{2}, \frac{1}{2}, 0, 0, \frac{1}{2})$
which is particular to these non-planar integrals.

The contribution of the all-hard regions can be calculated
by using the massless integral family and
IBP-reduction
[cf. the text below Eq.~\eqref{one-r1-8}].
Therefore we do not need a template integral for Region~1,
and we show the calculation of Region~1 separately.
The template integrals for Regions~2 to 14 are
calculated in the next subsection,
and can be found in the ancillary file to this paper~\cite{anci}.

\subsubsection*{Region 1 $(0,0,0,0,0,0,0)$ (all-hard region)}

As shown in Subsection~\ref{ss:one-top},
the contribution from the all-hard region 
can be expressed in terms of massless integrals
of the same topology.
The massless integral that is relevant here has been 
calculated in Ref.~\cite{Tausk:1999vh}.

%%%%%%%%%%%%%%%%%%%%%%%%%%%%%%%%%%%%%%%%%%%%%%%%%%%%%%%
\subsection{Template Integrals for Regions 2 to 14}
\label{ss:two-temp}

The template integral of each region is
expressed as
\begin{align}
\mathcal{T}^{(j)}_{\delta_1,\delta_2,\delta_3,\delta_4,\delta_5,\delta_6,\delta_7,\epsilon}&=
\int\! \mathfrak{D}^7\alpha^\delta~
\left( \mathcal{U}^{(j)} \right)^{-d/2}
e^{ -\mathcal{F}^{(j)}/\mathcal{U}^{(j)} }
\label{two-temp}
\end{align}
where $2\leq j\leq 14$.
For simplicity,
we omit the subscripts of 
$\mathcal{T}^{(j)}_{\delta_1,\delta_2,\delta_3,\delta_4,\delta_5,\delta_6,\delta_7,\epsilon}$
and write $\mathcal{T}^{(j)}$ when they are in the ordinary order.

\subsubsection*{Region 2 
$(0,\frac{1}{2},\frac{1}{2},0,0,\frac{1}{2},1)$, 
Region 3
$(\frac{1}{2},0,0,\frac{1}{2},1,\frac{1}{2},0)$} 

The Symanzik polynomials of Region~2 are
given by
\begin{align}
&\mathcal{U}^{(2)}=\alpha_1\alpha_{45}+\alpha_4\alpha_5
\label{two-r2-1}
\\
&\mathcal{F}^{(2)}=m_t^2\alpha_{45}~\mathcal{U}^{(2)}
+S(\alpha_2\alpha_3\alpha_5+\alpha_1\alpha_{45}\alpha_7+\alpha_4\alpha_5\alpha_7)
+T\alpha_1\alpha_3\alpha_6
+U\alpha_2\alpha_4\alpha_6
\,.
\label{two-r2-2}
\end{align}
The integration in $\alpha_7$ can be performed using the relation~\eqref{form1}.
Then, we perform the following change of variables
\begin{align}
\alpha_2\to \beta_1\beta_3/\beta_2,\quad
\alpha_3\to \beta_1\beta_2/\beta_3,\quad
\alpha_6\to \beta_2\beta_3/\beta_1
\label{two-r2-3}
\,,
\end{align}
and the template integral becomes 
\begin{align}
\mathcal{T}^{(2)}
&=
4S^{-1-\delta_7}
\int_{0}^{\infty}
\! \mathrm{d}\alpha_1\mathrm{d}\alpha_4\mathrm{d}\alpha_5
\mathrm{d}\beta_1\mathrm{d}\beta_2\mathrm{d}\beta_3~
\alpha_1^{\delta_1}
\alpha_4^{\delta_4}
\alpha_5^{\delta_5}
\beta_1^{\delta_{23\bar6}}
\beta_2^{\delta_{\bar236}}
\beta_3^{\delta_{2\bar36}}
\nonumber\\
&
\times
\frac{
\left( \alpha_1\alpha_4+\alpha_1\alpha_5+\alpha_4\alpha_5 \right)^{-d/2}
e^{-m_t^2\alpha_{45}-\left( S b_1^2\alpha_5+Tb_2^2\alpha_1+Ub_3^2\alpha_4\right)/\mathcal{U}^{(2)}}
}
{\Gamma [
\delta_1+1,
\delta_2+1,
\delta_3+1,
\delta_4+1,
\delta_5+1,
\delta_6+1
]}
\,.
\label{two-r2-4}
\end{align}
The integration in $\beta_1,\beta_2,\beta_3$ is now straightforward.
To integrate over the remaining variables,
it is easiest to integrate in $\alpha_1$ first using the relation~\eqref{form2}
and then integrate in $\alpha_4,\alpha_5$.
Finally we obtain the template integral as 
\begin{align}
&\mathcal{T}^{(2)}
=
\frac{1}{2}
(m_t^2)^{-(1+4\epsilon+\delta_{112344556})/2}
S^{-(3+\delta_{23\bar 677})/2}
T^{-(1+\delta_{\bar 2 36})/2}
U^{-(1+\delta_{2\bar 36})/2}
\nonumber\\
&\qquad \times
\frac{\Gamma [
\delta_{\bar0\bar1\bar2},
\frac{1+\delta_{23\bar 6}}{2},
\frac{1+\delta_{2\bar 36}}{2},
\frac{1+\delta_{\bar 236}}{2},
\frac{1+\delta_{112\bar 3\bar 6}}{2},
\frac{1+\delta_{00112344\bar 6}}{2},
\frac{1+\delta_{00112\bar 3556}}{2},
\frac{1+\delta_{0000112344556}}{2}
]}
{\Gamma [
\delta_1+1,
\delta_2+1,
\delta_3+1,
\delta_4+1,
\delta_5+1,
\delta_6+1,
\delta_{0011245}+1,
\frac{1-\delta_{00236}}{2}
]}
\,.
\label{two-r2-6}
\end{align}

The template integral of the Region~3 is
obtained by the replacement 
$\alpha_1\leftrightarrow\alpha_2$,
$\alpha_3\leftrightarrow\alpha_4$,
$\alpha_5\leftrightarrow\alpha_7$,
$T\leftrightarrow U$
of $\mathcal{T}^{(2)}$
\begin{align}
\mathcal{T}^{(3)}_{\delta_1,\delta_2,\delta_3,\delta_4,\delta_5,\delta_6,\delta_7,\epsilon}=
\mathcal{T}^{(2)}_{\delta_2,\delta_1,\delta_4,\delta_3,\delta_7,\delta_6,\delta_5,\epsilon}
\Big|_{T\leftrightarrow U}
\label{two-r2-7}
\,.
\end{align}

\subsubsection*{Region 4 $(0,0,0,1,1,1,0)$,
Region 5 $(0,0,1,0,0,1,1)$}

The Symanzik polynomials of Region~4 are given by
\begin{align}
&\mathcal{U}^{(4)}=\alpha_{12}\alpha_{37}+\alpha_{3}\alpha_{7}
\label{two-r4-1}
\\
&\mathcal{F}^{(4)}=m_t^2\alpha_{37}~\mathcal{U}^{(4)}
+S(\alpha_3\alpha_5\alpha_7+\alpha_1\alpha_{45}\alpha_7+\alpha_2\alpha_5\alpha_{37})
+T\alpha_1\alpha_3\alpha_6
\,.
\label{two-r4-2}
\end{align}
The integrations in $\alpha_4,\alpha_5,\alpha_6$
can be done
using the relation~\eqref{form1}.
Then the template integral is
\begin{align}
\mathcal{T}^{(4)}
&=
\frac{1}
{S^{2+\delta_{45}}T^{1+\delta_6}}
\int_{0}^{\infty}
\! \mathrm{d}\alpha_1\mathrm{d}\alpha_2\mathrm{d}\alpha_3\mathrm{d}\alpha_7~
\alpha_1^{-2+\delta_{1\bar4\bar6}}
\alpha_2^{\delta_2}
\alpha_3^{-1+\delta_{3\bar6}}
\alpha_7^{-1+\delta_{\bar47}}
\nonumber\\
&
\times
\frac{
\left(\alpha_{12}\alpha_{37}+\alpha_{3}\alpha_{7} \right)^{3-d/2+\delta_{456}}
\left(\alpha_{13}\alpha_{7}+\alpha_{2}\alpha_{37} \right)^{-1-\delta_{5}}
e^{-m_t^2\alpha_{37}}
}
{\Gamma [
\delta_1+1,
\delta_2+1,
\delta_3+1,
\delta_7+1
]}
\,.
\label{two-r4-3}
\end{align}
We introduce a Mellin-Barnes integral 
to separate
$\alpha_{12}\alpha_{37}+\alpha_{3}\alpha_{7}$
into 
two factors,
$\alpha_{13}\alpha_{7}+\alpha_{2}\alpha_{37}$
and 
$\alpha_1\alpha_3$,
then the integration in $\alpha_1$ and $\alpha_2$
can be done using the relation~\eqref{form2}.
The remaining integration is also
straightforward and we obtain
\begin{align}
\mathcal{T}^{(4)}=&
\int\mathrm{d}z_1
\frac{\Gamma[\delta_{\bar{0}\bar{1}\bar{2}},\delta_{001237},z_{\bar{1}}+\delta_{0267}+1,z_{\bar{1}},z_{1}+\delta_{0123\bar{6}},z_{1}+\delta_{1\bar{4}\bar{6}}-1,z_{1}+\delta_{\bar{0}\bar{4}\bar{5}\bar{6}}-1]}
{(m_t^2)^{\delta_{001237}}S^{2+\delta_{45}}T^{1+\delta_{6}}
\Gamma[\delta_{1}+1,\delta_{3}+1,\delta_{\bar{0}\bar{4}\bar{5}\bar{6}}-1,\delta_{7}+1,\delta_{0012237}+1,z_{1}+\delta_{\bar{0}\bar{4}\bar{6}}]}\,.
\label{template-npl4}
\end{align}

The template integral of Region~5
can be obtained in a similar manner
and the result is
\begin{align}
\mathcal{T}^{(5)}=&
\int\mathrm{d}z_1
\frac{\Gamma[\delta_{\bar{0}\bar{1}\bar{2}},\delta_{001245},z_{\bar{1}}+\delta_{0156}+1,z_{\bar{1}},z_{1}+\delta_{2\bar{3}\bar{6}}-1,z_{1}+\delta_{0124\bar{6}},z_{1}+\delta_{\bar{0}\bar{3}\bar{6}\bar{7}}-1]}
{(m_t^2)^{\delta_{001245}}S^{2+\delta_{37}}U^{1+\delta_{6}}
\Gamma[\delta_{2}+1,\delta_{4}+1,\delta_{5}+1,\delta_{0011245}+1,\delta_{\bar{0}\bar{3}\bar{6}\bar{7}}-1,z_{1}+\delta_{\bar{0}\bar{3}\bar{6}}]}\,.
\label{template-npl5}
\end{align}

\subsubsection*{Region 6 $(0,1,0,0,0,1,1)$,
Region 8 $(1,0,0,0,1,1,0)$}

The Symanzik polynomials of Region~6 are given by
\begin{align}
&\mathcal{U}^{(6)}=\alpha_{1}\alpha_{345}+\alpha_{34}\alpha_5
\label{two-r6-1}
\\
&\mathcal{F}^{(6)}=m_t^2\alpha_{345}~\mathcal{U}^{(6)}
+S(\alpha_2\alpha_3\alpha_5+\alpha_{34}\alpha_{5}\alpha_7+\alpha_1\alpha_{45}\alpha_{7})
+T\alpha_1\alpha_3\alpha_6
\,.
\label{two-r6-2}
\end{align}
The integrations in $\alpha_2,\alpha_6,\alpha_7$
are straightforward
using the relation~\eqref{form1}
and the template integral is given by
\begin{align}
\mathcal{T}^{(6)}
&=
\frac{1}
{S^{2+\delta_{27}}T^{1+\delta_6}}
\int_{0}^{\infty}
\! \mathrm{d}\alpha_1\mathrm{d}\alpha_3\mathrm{d}\alpha_4\mathrm{d}\alpha_5~
\alpha_1^{-1+\delta_{1\bar6}}
\alpha_3^{-2+\delta_{\bar23\bar6}}
\alpha_4^{\delta_4}
\alpha_5^{-1+\delta_{\bar25}}
\nonumber\\
&
\times
\frac{
\left(\alpha_{1}\alpha_{345}+\alpha_{34}\alpha_5 \right)^{3-d/2+\delta_{267}}
\left(\alpha_{34}\alpha_{5}+\alpha_{1}\alpha_{45} \right)^{-1-\delta_{7}}
e^{-m_t^2\alpha_{345}}
}
{\Gamma [
\delta_1+1,
\delta_3+1,
\delta_4+1,
\delta_5+1
]}
\,.
\label{two-r6-3}
\end{align}
This looks similar to Eq.~\eqref{two-r4-3}
but has one extra massive integral,
and thus it is necessary to
introduce two Mellin-Barnes integrals,
giving
\begin{align}
\mathcal{T}^{(6)}=&
\int\mathrm{d}z_1\mathrm{d}z_2
\frac{\Gamma[\delta_{015},z_{\bar{1}},z_{1}+\delta_{1\bar{6}},z_{1}+\delta_{\bar{0}\bar{2}\bar{6}\bar{7}}-1,z_{\bar{2}}+\delta_{0124}+1]}
{(m_t^2)^{\delta_{001345}}S^{2+\delta_{27}}T^{1+\delta_{6}}
\Gamma[\delta_{1}+1,\delta_{3}+1,\delta_{4}+1,\delta_{5}+1]}\nonumber\\
&\qquad\times
\frac{\Gamma[z_{\bar{1}\bar{2}}+\delta_{0012456}+1,z_{\bar{2}},z_{2}+\delta_{\bar{0}\bar{1}\bar{2}},z_{12}+\delta_{\bar{2}3\bar{6}}-1]}
{\Gamma[\delta_{\bar{0}\bar{2}\bar{6}\bar{7}}-1,z_{1}+\delta_{\bar{0}\bar{2}\bar{6}},z_{\bar{2}}+\delta_{0011245}+1]}\,.
\label{template-npl6}
\end{align}

The template integral of Region~8
can be obtained in a similar manner
and the result is
\begin{align}
\mathcal{T}^{(8)}=&
\int\mathrm{d}z_1\mathrm{d}z_2
\frac{\Gamma[\delta_{\bar{0}\bar{1}\bar{2}},z_{\bar{1}},z_{1}+\delta_{\bar{1}4\bar{6}}-1,z_{1}+\delta_{\bar{0}\bar{1}\bar{5}\bar{6}}-1,z_{\bar{1}\bar{2}}+\delta_{067}]}
{(m_t^2)^{\delta_{002347}}S^{2+\delta_{15}}U^{1+\delta_{6}}
\Gamma[\delta_{2}+1,\delta_{3}+1,\delta_{4}+1,\delta_{\bar{0}\bar{1}\bar{5}\bar{6}}-1]}\nonumber\\
&\qquad\times
\frac{\Gamma[z_{\bar{2}},z_{2}+\delta_{3}+1,z_{12}+\delta_{2\bar{6}},z_{12}+\delta_{0234\bar{6}}]}
{\Gamma[\delta_{7}+1,z_{1}+\delta_{\bar{0}\bar{1}\bar{6}},z_{12}+\delta_{\bar{1}34\bar{6}}]}\,.
\label{template-npl8}
\end{align}

\subsubsection*{Region 7 $(0,1,1,0,0,0,1)$,
Region 9 $(1,0,0,1,1,0,0)$}

The Symanzik polynomials of Region~7 are given by
\begin{align}
&\mathcal{U}^{(7)}=\alpha_{1}\alpha_{456}+\alpha_{4}\alpha_{56}
\label{two-r7-1}
\\
&\mathcal{F}^{(7)}=m_t^2\alpha_{456}~\mathcal{U}^{(7)}
+S(\alpha_{4}\alpha_{5}\alpha_7+\alpha_1\alpha_{45}\alpha_{7})
+T\alpha_1\alpha_3\alpha_6
+U\alpha_2\alpha_4\alpha_6
\,.
\label{two-r7-2}
\end{align}
The template integral is obtained in a similar way
as that of Region~6 and the resulting 
two-dimensional integral is given by
\begin{align}
\mathcal{T}^{(7)}=&
\int\mathrm{d}z_1\mathrm{d}z_2
\frac{\Gamma[\delta_{014},z_{\bar{1}}+\delta_{0012345}+1,z_{1}+\delta_{\bar{2}\bar{3}6}-1,z_{1}+\delta_{\bar{0}\bar{2}\bar{3}\bar{7}}-1,z_{\bar{1}\bar{2}}+\delta_{\bar{1}3}]}
{(m_t^2)^{\delta_{001456}}S^{1+\delta_{7}}T^{1+\delta_{3}}U^{1+\delta_{2}}
\Gamma[\delta_{1}+1,\delta_{4}+1,\delta_{5}+1,\delta_{6}+1]}\nonumber\\
&\qquad\times
\frac{\Gamma[z_{\bar{1}\bar{2}}+\delta_{0235}+1,z_{\bar{2}},z_{2}+\delta_{1\bar{3}},z_{12}+\delta_{\bar{0}\bar{2}\bar{3}}]}
{\Gamma[\delta_{\bar{0}\bar{2}\bar{3}\bar{7}}-1,z_{1}+\delta_{\bar{0}\bar{2}\bar{3}},z_{\bar{1}\bar{2}}+\delta_{0012345}+1]}\,.
\label{template-npl7}
\end{align}

The template integral of Region~9
can also be obtained in a similar manner
and the result is given by
\begin{align}
\mathcal{T}^{(9)}=&
\int\mathrm{d}z_1\mathrm{d}z_2
\frac{\Gamma[\delta_{023},z_{\bar{1}}+\delta_{0012347}+1,z_{1}+\delta_{\bar{0}\bar{1}\bar{4}\bar{5}}-1,z_{1}+\delta_{\bar{1}\bar{4}6}-1,z_{\bar{1}\bar{2}}+\delta_{\bar{2}4}]}
{(m_t^2)^{\delta_{002367}}S^{1+\delta_{5}}T^{1+\delta_{1}}U^{1+\delta_{4}}
\Gamma[\delta_{2}+1,\delta_{3}+1,\delta_{\bar{0}\bar{1}\bar{4}\bar{5}}-1,\delta_{6}+1]}\nonumber\\
&\qquad\times
\frac{\Gamma[z_{\bar{1}\bar{2}}+\delta_{0147}+1,z_{\bar{2}},z_{2}+\delta_{2\bar{4}},z_{12}+\delta_{\bar{0}\bar{1}\bar{4}}]}
{\Gamma[\delta_{7}+1,z_{1}+\delta_{\bar{0}\bar{1}\bar{4}},z_{\bar{1}\bar{2}}+\delta_{0012347}+1]}\,.
\label{template-npl9}
\end{align}

\subsubsection*{Region 10 $(1,1,0,0,0,0,1)$,
Region 11 $(1,1,0,0,1,0,0)$}

The Symanzik polynomials of Region~10 are given by
\begin{align}
&\mathcal{U}^{(10)}=\alpha_{34}\alpha_{56}
\label{two-r10-1}
\\
&\mathcal{F}^{(10)}=m_t^2\alpha_{3456}~\mathcal{U}^{(10)}
+S(\alpha_{2}\alpha_{3}\alpha_5+\alpha_{34}\alpha_{5}\alpha_{7})
+T\alpha_1\alpha_3\alpha_6
+U\alpha_2\alpha_4\alpha_6
\,.
\label{two-r10-2}
\end{align}
The template integral is obtained in a similar way
as that of Region~5 and the resulting 
one-dimensional integral is given by
\begin{align}
\mathcal{T}^{(10)}=&
\int\mathrm{d}z_1
\frac{\Gamma[\delta_{034},\delta_{056},z_{\bar{1}}+\delta_{\bar{1}\bar{2}3}-1,z_{\bar{1}}+\delta_{\bar{2}5\bar{7}}-1,z_{\bar{1}},z_{1}+\delta_{2}+1,z_{1}+\delta_{4}+1,z_{1}+\delta_{\bar{1}6}]}
{(m_t^2)^{\delta_{003456}}S^{2+z_{1}+\delta_{27}}T^{1+\delta_{1}}U^{z_{\bar{1}}}
\Gamma[\delta_{2}+1,\delta_{3}+1,\delta_{4}+1,\delta_{\bar{1}\bar{2}34},\delta_{5}+1,\delta_{6}+1,\delta_{\bar{1}\bar{2}56\bar{7}}-1]}\,.
\label{template-npl10}
\end{align}

The template integral of Region~11
can be obtained in a similar manner
and the result is given by
\begin{align}
\mathcal{T}^{(11)}=&
\int\mathrm{d}z_1
\frac{\Gamma[\delta_{034},\delta_{067},z_{\bar{1}}+\delta_{\bar{1}\bar{2}4}-1,z_{\bar{1}}+\delta_{\bar{1}\bar{5}7}-1,z_{\bar{1}},z_{1}+\delta_{1}+1,z_{1}+\delta_{3}+1,z_{1}+\delta_{\bar{2}6}]}
{(m_t^2)^{\delta_{003467}}S^{2+z_{1}+\delta_{15}}T^{z_{\bar{1}}}U^{1+\delta_{2}}
\Gamma[\delta_{1}+1,\delta_{3}+1,\delta_{4}+1,\delta_{\bar{1}\bar{2}34},\delta_{6}+1,\delta_{7}+1,\delta_{\bar{1}\bar{2}\bar{5}67}-1]}\,.
\label{template-npl11}
\end{align}

\subsubsection*{Region 12 $(1,1,0,0,1,1,1)$}

The Symanzik polynomials of Region~12 are given by
\begin{align}
&\mathcal{U}^{(12)}=\alpha_{34}\alpha_{12567}
\label{two-r12-1}
\\
&\mathcal{F}^{(12)}=m_t^2\alpha_{34}~\mathcal{U}^{(12)}
+S(\alpha_{2}\alpha_{3}\alpha_5+\alpha_{1}\alpha_{4}\alpha_7+\alpha_{34}\alpha_{5}\alpha_{7})
+T\alpha_1\alpha_3\alpha_6
+U\alpha_2\alpha_4\alpha_6
\,.
\label{two-r12-2}
\end{align}
We introduce Mellin-Barnes integrals
four times,
and the template integral is given by
\begin{align}
\mathcal{T}^{(12)}=&
\int\mathrm{d}z_1\mathrm{d}z_2\mathrm{d}z_3\mathrm{d}z_4
\frac{\Gamma[\delta_{034},z_{\bar{1}},z_{\bar{2}},z_{12}+\delta_{6}+1,z_{\bar{1}\bar{2}\bar{3}}+\delta_{\bar{0}\bar{1}\bar{2}\bar{5}\bar{6}}-2,z_{\bar{3}},z_{13}+\delta_{3}+1,z_{23}+\delta_{2}+1]}
{(m_t^2)^{\delta_{034}}S^{3+z_{12}+\delta_{012567}}T^{z_{\bar{1}}}U^{z_{\bar{2}}}
\Gamma[\delta_{1}+1,\delta_{2}+1,\delta_{3}+1,\delta_{4}+1]}\nonumber\\
&\qquad\times
\frac{\Gamma[z_{\bar{1}\bar{3}\bar{4}}+\delta_{\bar{0}\bar{1}\bar{2}4\bar{5}\bar{6}\bar{7}}-2,z_{\bar{2}\bar{3}\bar{4}}+\delta_{\bar{0}\bar{2}\bar{5}\bar{6}\bar{7}}-2,z_{\bar{4}},z_{34}+\delta_{5}+1,z_{1234}+\delta_{012567}+3]}
{\Gamma[\delta_{5}+1,\delta_{6}+1,\delta_{\bar{0}\bar{0}\bar{1}\bar{2}\bar{5}\bar{6}\bar{7}}-1,\delta_{7}+1,z_{\bar{4}}+\delta_{\bar{0}\bar{1}\bar{2}34\bar{5}\bar{6}\bar{7}}-1]}\,.
\label{template-npl12}
\end{align}

\subsubsection*{Region 13 $(1,1,1,1,0,0,1)$, Region 14 $(1,1,1,1,1,0,0)$}

The Symanzik polynomials of Region~13 are given by
\begin{align}
&\mathcal{U}^{(13)}=\alpha_{1234}\alpha_{56}
\label{two-r13-1}
\\
&\mathcal{F}^{(13)}=m_t^2\alpha_{56}\mathcal{U}^{(13)}
+S(\alpha_{134}\alpha_{5}\alpha_7+\alpha_{2}\alpha_{37}\alpha_5)
+T\alpha_1\alpha_3\alpha_6
+U\alpha_2\alpha_4\alpha_6
\label{two-r13-2}
\,.
\end{align}
It is necessary to introduce the Mellin-Barnes integral twice
in order to separate the terms proportional to $S,T$, or $U$, respectively,
and we obtain 
\begin{align}
\mathcal{T}^{(13)}=&
\int\mathrm{d}z_1\mathrm{d}z_2
\frac{\Gamma[\delta_{056},z_{\bar{1}}+\delta_{\bar{0}\bar{1}\bar{2}\bar{3}\bar{4}6}-1,z_{\bar{1}},z_{1}+\delta_{5\bar{7}},z_{\bar{2}}+\delta_{\bar{0}\bar{1}\bar{3}\bar{4}}-1,z_{\bar{1}\bar{2}}+\delta_{\bar{0}\bar{1}\bar{2}\bar{3}}-1,z_{\bar{2}}]}
{(m_t^2)^{\delta_{056}}S^{1+z_{\bar{1}}+\delta_{7}}T^{z_{\bar{2}}}U^{2+z_{12}+\delta_{01234}}
\Gamma[\delta_{1}+1,\delta_{2}+1,\delta_{3}+1,\delta_{\bar{0}\bar{0}\bar{1}\bar{2}\bar{3}\bar{4}}]}\nonumber\\
&\qquad\times
\frac{\Gamma[z_{2}+\delta_{1}+1,z_{12}+\delta_{3}+1,z_{12}+\delta_{01234}+2]}
{\Gamma[\delta_{4}+1,\delta_{5}+1,\delta_{6}+1,\delta_{\bar{0}\bar{1}\bar{2}\bar{3}\bar{4}56\bar{7}}-1]}\,.
\label{template-npl13}
\end{align}

The template integral of Region~14
can be obtained in a similar manner
and the result is
\begin{align}
\mathcal{T}^{(14)}=&
\int\mathrm{d}z_1\mathrm{d}z_2
\frac{\Gamma[\delta_{067},z_{\bar{1}}+\delta_{\bar{0}\bar{1}\bar{2}\bar{3}\bar{4}6}-1,z_{\bar{1}},z_{1}+\delta_{\bar{5}7},z_{\bar{2}}+\delta_{\bar{0}\bar{2}\bar{3}\bar{4}}-1,z_{\bar{1}\bar{2}}+\delta_{\bar{0}\bar{1}\bar{2}\bar{4}}-1,z_{\bar{2}}]}
{(m_t^2)^{\delta_{067}}S^{1+z_{\bar{1}}+\delta_{5}}T^{2+z_{12}+\delta_{01234}}U^{z_{\bar{2}}}
\Gamma[\delta_{1}+1,\delta_{2}+1,\delta_{3}+1,\delta_{\bar{0}\bar{0}\bar{1}\bar{2}\bar{3}\bar{4}}]}\nonumber\\
&\qquad\times
\frac{\Gamma[z_{2}+\delta_{2}+1,z_{12}+\delta_{4}+1,z_{12}+\delta_{01234}+2]}
{\Gamma[\delta_{4}+1,\delta_{6}+1,\delta_{7}+1,\delta_{\bar{0}\bar{1}\bar{2}\bar{3}\bar{4}\bar{5}67}-1]}\,.
\label{template-npl14}
\end{align}

In the limit where all the regularization parameters go to zero,
Eq.~\eqref{template-npl13} becomes
\begin{align}
\mathcal{T}^{(13)}=
&-2\epsilon
\int \mathrm{d}z_1\mathrm{d}z_2
\frac{S^{-1+z_1}T^{z_2}}{U^{2+z_1+z_2}}
\Gamma
[
z_{\bar1}-1,z_{\bar1},z_1,z_{\bar2}-1,z_{\bar2},z_2+1,z_{\bar1\bar2}-1,
z_{12}+1,z_{12}+2
]
\nonumber\\
&+\mbox{(one-dimensional Mellin-Barnes integrals)}
+\mathcal{O}(\epsilon^2)
\label{two-r13-4}
\,,
\end{align}
which means that the two-dimensional integral
does not contribute at $\epsilon^0$-order.
Thus, the representation~\eqref{template-npl13}
is most useful when the required order is $\epsilon^0$.

There is another representation of $\mathcal{T}^{(13)}$
which is more suitable if we require calculation beyond order $\epsilon^0$:
\begin{align}
\mathcal{T}^{(13)}=&
\int\mathrm{d}z_1
\frac{\Gamma[\delta_{\bar{0}},\delta_{\bar{0}\bar{1}\bar{2}},\delta_{012}+1,\delta_{056},\delta_{5\bar{7}},z_{\bar{1}}+\delta_{\bar{0}\bar{1}}-1,z_{\bar{1}},z_{1}+1,z_{1}+\delta_{\bar{0}\bar{1}\bar{2}6}]}
{(m_t^2)^{\delta_{056}}S^{2+z_{1}+\delta_{7}}T^{z_{\bar{1}}}U^{1+\delta_{012}}
\Gamma[\delta_{\bar{0}\bar{0}\bar{1}\bar{2}},\delta_{2}+1,\delta_{5}+1,\delta_{6}+1,z_{\bar{1}}+\delta_{\bar{0}},z_{1}+\delta_{\bar{0}\bar{1}\bar{2}56\bar{7}}]}\nonumber\\
&
-\int\mathrm{d}z_1\mathrm{d}z_2
\frac{\Gamma[\delta_{\bar{0}},\delta_{\bar{0}\bar{1}\bar{2}},\delta_{056},z_{\bar{1}}+\delta_{\bar{0}\bar{1}\bar{2}5\bar{7}}-1,z_{\bar{1}},z_{1}+\delta_{1}+1,z_{1}+\delta_{012}+1]}
{(m_t^2)^{\delta_{056}}S^{2+z_{1\bar{2}}+\delta_{0127}}T^{1+z_{\bar{1}2}}
\Gamma[\delta_{1}+1,\delta_{\bar{0}\bar{0}\bar{1}\bar{2}},\delta_{2}+1,\delta_{5}+1]}\nonumber\\
&\qquad\times
\frac{\Gamma[z_{1\bar{2}}+\delta_{6},z_{\bar{2}},z_{2}+1,z_{\bar{1}2}+\delta_{\bar{0}\bar{1}}]}
{\Gamma[\delta_{6}+1,z_{\bar{2}}+\delta_{\bar{0}\bar{1}\bar{2}56\bar{7}}-1,z_{2}+\delta_{\bar{0}}+1]}\,.
\label{template-npl13-b}
\end{align}

This representation consists of
two integrals with at most two kinematic parameters,
and their calculation is simpler.
It has some disadvantages, however;
it holds only when $\delta_3$ and $\delta_4$
are non-negative integers,
and Eq.~\eqref{template-npl13-b} is shown in the 
special case that
$\delta_3=0, \delta_4=0$,
since that is the typical situation
(here we may set these values because $\mathcal{T}^{(13)}$
is regular in $\delta_3$ and $\delta_4$).
Another disadvantage is that each integral
produces singularities in $\epsilon$ which cancel
in their sum,
however
as a by-product,
higher-order derivatives of $\Gamma$-functions
contribute to the $\epsilon^0$ order.

%%%%%%%%%%%%%%%%%%%%%%%%%%%%%%%%%%%%%%%%%%%%%%%%%%%%%%%
\subsection{Analytic Continuation}
\label{ss:anacon}

As mentioned in the text below Eq.~\eqref{mb1},
the Mellin-Barnes integrals in the template integrals
are assumed to be regularized
by choosing suitable values of $\delta_j$.
However, the quantity we need is the one
where $\delta_j\to0$ for all $j$.
Therefore we need to analytically continue 
the Mellin-Barnes integrals
in terms of $\delta_j$.
We describe the procedure of the analytic continuation
showing the case of Region~7, which is one of the most involved cases,
as a concrete example.\footnote{
In terms of the dimension of the Mellin-Barnes integrals,
the most involved is Region~12.
However, the analytic continuation 
turns out to be rather simple in this region.
}
We take the limit of the ascending order of $\delta_j$,
\begin{align}
\lim_{\epsilon,\delta_7,\delta_6,\delta_5,\delta_4,\delta_3,\delta_2,\delta_1\to0}
\mathcal{T}^{(7)}
\,.
\label{ana1}
\end{align}

As mentioned below Eq.~\eqref{mb1},
in general 
the integral contours of $z_1,...,z_4$
are assumed to be straight lines parallel to the imaginary axis
[cf. Fig.~\ref{fig:contour1} (b)].
For Region~7, we have only $z_1, z_2$ as integration variables.
We choose $\mathrm{Re}(z_1)=-1/5$, $\mathrm{Re}(z_2)=-1/3$.
%$\mathrm{Re}(z_3)=-4/7$, $\mathrm{Re}(z_4)=-3/11$.
Then, we may choose 
\begin{align}
\delta_1=0,~
\delta_2=0,~
\delta_3=-\frac{13}{30},~
\delta_4=\frac{1}{300},~
\delta_5=\frac{1}{500},~
\delta_6=\frac{17881}{15540},~
\delta_7=-\frac{64003}{62160},~
\epsilon=-\frac{2507}{10360}
\label{ana2}
\end{align}
to regularize the template integral $\mathcal{T}^{(7)}$.\footnote{
These values can be changed,
provided they do not
cross the integration contours.
}
We try to set as many parameters as possible to zero
in ascending $j$ order in Eq.~\eqref{ana2}.
The limit of $\delta_{1,2}\to 0$ in Eq.~\eqref{ana1}
is now trivial.

The analytic continuation of $\delta_3$ from $-13/30$ to 0
makes the first rightmost left poles of 
$\Gamma (z_1+\delta_{\bar 36}-1)$,
$\Gamma (z_2-\delta_{3})$,
and
$\Gamma (z_{12}-\delta_{03})$
into right poles,
which must be compensated by adding their residues.
This analytic continuation procedure is automatized in the \texttt{Mathematica} package 
\texttt{MB.m}~\cite{Czakon:2005rk}.
After the analytic continuation in terms of $\delta_3$,
the integral depends on $\delta_4,...,\delta_7$ and $\epsilon$,
and we repeat the same procedure for $\delta_4$, then, $\delta_5$ and so on.
In this way,
we obtain a combination of integrals
for which the arguments of the $\Gamma$-functions in the integrand
contain only $z_1$ and $z_2$ such as
\begin{align}
\int \mathrm{d}z_1\mathrm{d}z_2
~
\frac{
\Gamma[
1-z_1,-1+z_1,-1+z_1,-z_{12},-z_2,z_2,z_{12}]
}
{\Gamma ( z_1)}
\label{ana4}
\,.
\end{align}
The methods to solve these integrals are
explained in the next subsection.

%%%%%%%%%%%%%%%%%%%%%%%%%%%%%%%%%%%%%%%%%%%%%%%%%%%%%%%
\subsection{Solving the Mellin-Barnes Integrals}
\label{ss:mb}

The usual idea to solve the Mellin-Barnes integral 
is to apply the first and the second Barnes lemma
and variants of them.
The \texttt{Mathematica} package
\texttt{barnesroutines.m}~\cite{hepforge}
performs this procedure in an automatic way,
and solves some of the Mellin-Barnes integrals
we encounter.
Unfortunately,
not all of them are solved by this package,
and we describe here how to treat such cases.
The essential points are mentioned in Ref.~\cite{Davies:2018ood},
and we fit or extend them to our integrals.

\subsubsection*{Three- and Four-Dimensional Mellin-Barnes Integrals}

The template integral of the Region~12~\eqref{template-npl12}
is expressed as a four-dimensional Mellin-Barnes integral.
Thus, the contribution from Region~12
contain a four-dimensional integral
of the form
\begin{align}
\int \mathrm{d}z_1\mathrm{d}z_2\mathrm{d}z_3\mathrm{d}z_4
&\left(\frac{T}{S}\right)^{z_1}
\left(\frac{U}{S}\right)^{z_2}
\frac{
\Gamma [-z_4,
1+z_{34},
-2-z_{134},
-2-z_{234},
3+z_{1234}
]
}
{\Gamma (-1-z_4)}
\nonumber\\
&\times
\Gamma[
-z_1,-z_2,-z_3,
1+z_{12},
1+z_{13},
1+z_{23},
-2-z_{123}
]
\,.
\label{sol4-1}
\end{align}
We use the relation
\begin{align}
&
\frac{
\Gamma [-z_4,
1+z_{34},
-2-z_{134},
-2-z_{234},
3+z_{1234}
]
}
{\Gamma (-1-z_4)}
\nonumber\\
&=
\Gamma[
-2-z_{134},
-1-z_{234},
1+z_{34},
3+z_{1234}
]
\nonumber\\
&\quad +
\frac{
\Gamma[2+z_{23},
-2-z_{134},
-2-z_{234},
1+z_{34},
3+z_{1234}
]
}
{\Gamma(1+z_{23})}
\label{sol4-2}
\end{align}
to reduce the number of the $\Gamma$-functions whose argument contains $z_4$
from 6 to 4.
Now one can apply the first Barnes lemma to solve the $z_4$-integral.
The resulting  three-dimensional integral
can be easily reduced to a sum of two-dimensional integrals
since the $z_3$-integral can be solved by the variants of the first and second Barnes lemmas.
Thus, we have two-dimensional integrals,
and the way to solve them will be explained below.

Note that the reduction in Eq.~\eqref{sol4-2}
can be done only after the limits $\delta_j\to 0$ and $\epsilon \to0$
since the $\Gamma$-functions of the denominator and the numerator 
have different dependence on $\epsilon$,
thus the cancellation does not occur
before the limits have been taken.

The content of this subsection
is not formulated in an algorithmic way
and has been done manually.

\subsubsection*{Two-Dimensional Mellin-Barnes Integrals}

In the cases of integrals with no argument such as
\begin{align}
\int \mathrm{d}z_1\mathrm{d}z_2
~\Gamma[
-z_1,-1+z_1,-z_2,z_2,1-z_{12},-1+z_{12}]
\psi(z_1)\psi(1-z_{12})
\label{solmb2-1}
\,,
\end{align}
or integrals with a single argument of the form 
\begin{align}
\int \mathrm{d}z_1\mathrm{d}z_2
~X^{z_2}
\Gamma[
-z_1,-1+z_1,-z_2,z_2,1-z_{12},-1+z_{12}]
\psi(z_1)\psi(1-z_{12})
\label{solmb2-2}
\,,
\end{align}
we first reduce them to
a one-dimensional integral using the generalized
Barnes lemma~\cite{Davies:2018ood}
\begin{align}
&
\int_{C}^{}
\frac{dz}{2\pi i}
\frac{\Gamma [a_1-z,a_2-z,a_3+z,a_4+z,a_5+z]}{\Gamma (-a_6+z)}
\nonumber\\
&\hspace{2cm}=
\frac{\Gamma [a_{13},a_{23},a_{14},a_{24},a_{15},a_{25}]}
{\Gamma [a_{1235},a_{1245},-a_{56}]}
\, _3F_2
\left( 
\begin{array}[]{c}
	a_{15},a_{25},a_{123456}\\
	a_{1235},a_{1245}
\end{array};1
\right)
\,,
\label{solmb2-10}
\end{align}
where $_3F_2$ is the generalized hypergeometric function
~\cite{hypgeo1,hypgeo2}.
A useful corollary of Eq.~\eqref{solmb2-10} is presented in Appendix~\ref{app:mb}.
The resulting  one-dimensional integrals
can be solved by the method below.

Integrals with two different arguments are difficult to solve.
However
in our case, such integrals only appear at higher orders in $\epsilon$
so we do not need to consider them.

\subsubsection*{One-Dimensional Mellin-Barnes Integrals}

For integrals with no argument such as
\footnote{
After the analytic continuation described in Subsection~\ref{ss:anacon},
we may have an expression where some of the poles merge.
The following procedure can be applied also in these cases.
}
\begin{align}
\int \mathrm{d}z_1
~\Gamma[
-z_1,-1+z_1,1-z_1,-1+z_1]
\psi (z_1)\psi (-2-z_1)
\,,
\label{solmb1}
\end{align}
we evaluate them numerically using \texttt{Mathematica}
and apply the PSLQ algorithm~\cite{PSLQ1,PSLQ2}
to fit them 
to a basis of constants
which consists of all possible products of
\begin{align}
\{
1,\gamma_E,
\pi^2,\zeta_3
,\zeta_5
\}
\label{solmb2}
\end{align}
up to a transcendental weight of five.
The results turn out to 
only require constants to weight four.
Typically, 50--70 digits of the numerical result 
are sufficient to obtain the correct answer,
which we verify with 100 more digits.

The integrals with one argument 
typically have the form
\begin{align}
\int \mathrm{d}z_1
X^{z_1}
~\Gamma[
-z_1,-1+z_1,1-z_1,-1+z_1],
\qquad
X=\frac{X_1}{X_2},
\qquad 
X_1,X_2\in \{S,T,U\}
\,.
\label{solmb3}
\end{align}
Various combinations of $X_1, X_2$
appear since the template integrals
contain them.
We obtain the series expansion of Eq.~\eqref{solmb3}
by taking the residue of the left poles or the right poles.
By adjusting which poles we consider,
we can choose the series
in terms of either $T/S, U/S$, or $T/U$:
\footnote{
Below Eq.~\eqref{series},
it was stated that we use the normal equal sign for series representations
when the hierarchy is obvious.
However, here we have to introduce additional assumptions
for $X= T/S, U/S, T/U,$ since hierarchies between the positive Mandelstam variables 
have not been fixed up to this point.
Thus we use the sign ``$\myeq$" in Eq.~\eqref{solmb4},
indicating that a certain analytic continuation should be performed
in order to ensure $X\ll 1$.
}
\begin{align}
\int \mathrm{d}z_1
X^{z_1}
~\Gamma[
-z_1,-1+z_1,1-z_1,-1+z_1]
\myeq \sum_{n_1=0}^4 \sum_{n_2=0}^\infty 
(\log X)^{n_1}  X^{n_2}
\,,
\label{solmb4}
\end{align}
Now we apply analytic continuation
and obtain
\begin{align}
X=T/S:&\quad
\log X
\myeq 
h_0-\log s+i\pi,\qquad
X
\myeq -v
\label{solmb5}
\\
X=U/S:&\quad
\log X
\myeq 
-h_1-\log s+i\pi,\qquad
X\myeq -(1-v)
\label{solmb6}
\\
X=T/U:&\quad
\log X
\myeq 
h_0+h_1,\qquad
X\myeq \frac{v}{1-v}
\,.
\label{solmb7}
\end{align}
Recall that $h_1=-\log (1-v)$.
We fit the series with HPL and express the result
in terms of $h_N$.
In the case of~\eqref{solmb5},
the series in $v$ is directly fit to $h_N$.
In the case of~\eqref{solmb6},
we first fit the series to HPL with the argument of $(1-v)$
and then express them in terms of $h_N$.
In the case of~\eqref{solmb7},
we first fit the series to HPL with the argument of $v/(1-v)$
and then express them in terms of $h_N$.
Taking into account that
$0\leq v\leq 1$ and 
the brach cut of $h_{N>0}$ lies on the real axis of $v>1$,
we never cross the branch cut 
in the above procedure.
The information of the branch cut is encoded 
in the analytic continuation of $\log X$.

We already cover all of the combinations of $X=X_1/X_2$,
so the calculation of the crossed diagrams
can be done with the same procedure.
For our sample integral~\eqref{two-1},
there are about 50 one-dimensional Mellin-Barnes integrals
which are treated in this way.

%%%%%%%%%%%%%%%%%%%%%%%%%%%%%%%%%%%%%%%%%%%%%%%%%%%%%%%
\subsection{Combining the Results}
\label{ss:total}

Summing the contributions from all the relevant regions,
we obtain for our sample integral 
\begin{align}
I=&
%\frac{e^{2i\pi\epsilon}e^{-2\epsilon\gamma_E}}{s^{3+2\epsilon}}
%\left[ 
\frac{i\pi^3 e^{-2\epsilon\gamma_E}}{m_ts^{\frac{5}{2}}\sqrt{v}\sqrt{1-v}}
\left( \frac{1}{\epsilon}-2\log (m_t^2)-10\log 2\right)
%\frac{i\pi^3\sqrt{s}}{m_t\sqrt{v(1-v)}}
%\left[ \frac{1}{\epsilon}-2\log \left( \frac{m_t^2}{32s} \right) \right]
\nonumber\\
&+
\frac{e^{2i\pi\epsilon}e^{-2\epsilon\gamma_E}}{s^{3+2\epsilon}}
\sum_{i_1=-1}^0\sum_{i_2=0}^{4+i_1} \frac{d_{i_1,i_2}}{v(1-v)} \epsilon ^{i_1} \log ^{i_2} (m_t)
%\right]
+\mathcal{O}(m_t,\epsilon)
\end{align}
where
\begin{align}
d_{-1,3}&=-\frac{4}{3}\,,\qquad
d_{-1,2}=h_0 (4 v+2)+h_1 (4 v-6)+6 i \pi\,,\nonumber\\
d_{-1,1}&=-h_0^2+2 h_0 h_1+8 h_0 v+2 i \pi  h_0 (2 v-1)-h_1^2+8 h_1 (v-1)+2 i \pi  h_1 (2 v-1)-\frac{10 \pi ^2}{3}+8 i \pi\,,\nonumber\\
d_{-1,0}&=-\frac{1}{2} i \pi  h_0^2+4 h_0 h_1+i \pi  h_0 h_1-\frac{1}{2} i \pi  h_1^2-8 i \pi+4 i \pi  h_1 v\nonumber\\
&+\frac{1}{6} h_0^3 (1-2 v)+\frac{1}{3} \pi ^2 h_0 (5-8 v)+h_1 (8-8 v)+\pi ^2 h_1 \left(1-\frac{8 v}{3}\right)-\frac{4 i \pi ^3}{3}\nonumber\\
&+h_0^2 h_1 \left(\frac{1}{2}-v\right)+h_0 h_1^2 \left(\frac{1}{2}-v\right)+4 i \pi  h_0 (v-1)-8 h_0 v+\frac{1}{6} h_1^3 (1-2 v)\,,\nonumber\\
d_{0,4}&=-\frac{10}{3}\,,\qquad
d_{0,3}=h_0 (4-8 v)+h_1 (4-8 v)-\frac{20 i \pi }{3}\,,\nonumber\\
d_{0,2}&=-h_0^2+6 h_0 h_1-2 i \pi  h_0 (6 v+1)-h_1^2-2 i \pi  h_1 (6 v-7)+\frac{47 \pi ^2}{3}\,,\nonumber\\
d_{0,1}&=-i \pi  h_0^2+16 h_0 h_1-16 i \pi  h_0+16 i \pi  h_1-32 i \pi+\pi ^2 h_0 \left(2 v-\frac{7}{3}\right)\nonumber\\
&+\frac{1}{3} h_0^3 (-2 v-1)-2 i \pi  h_0 h_1-i \pi  h_1^2-i \pi ^3+16 \pi ^2+\pi ^2 h_1 \left(2 v+\frac{1}{3}\right)\nonumber\\
&+h_0^2 h_1 (-2 v-1)+h_0 h_1^2 (3-2 v)-32 h_0 v+h_1^3 \left(1-\frac{2 v}{3}\right)-32 h_1 (v-1)+6 \zeta_{3}\,,\nonumber\\
d_{0,0}&=8 h_0 h_1^2-32 h_0 h_1-12 i \pi  h_1 h_2\
+h_1 h_3 (8-16 v)+h_4 (60-48 v)-16 \pi ^2+\frac{22}{3} \pi ^2 h_2 (1-2 v)\nonumber\\
&+h_0^2 h_1 (8-8 v)+h_0^2 h_2 (10-8 v)+\frac{11}{12} \pi ^2 h_1^2 (1-4 v)
+h_0 h_1^3 \left(\frac{1}{2}-2 v\right)+\frac{1}{8} h_1^4 (5-4 v)\nonumber\\
&+h_0^2 h_1^2 \left(-v-\frac{1}{4}\right)+16 h_0 h_2 (v-2)-\frac{2}{3} \pi ^2 h_1 (v-13)
-16 i \pi  h_0 (v-2)-\frac{8}{3} h_1^3 (v-1)\nonumber\\
&-\frac{8 h_0^3 v}{3}-8 i \pi  h_0^2 (v-1)+8 i \pi  h_1^2 v
+8 i \pi  h_0 h_2 (v+1)-\frac{2}{3} \pi ^2 h_0 (v+12)-16 i \pi  h_1 (v+1)\nonumber\\
&-\frac{1}{6} i \pi  h_0^3 (2 v-11)+\frac{1}{6} i \pi ^3 h_1 (2 v-7)+\pi ^4 \left(\frac{151 v}{90}-\frac{19}{9}\right)
+h_0^3 h_1 \left(2 v-\frac{3}{2}\right)-8 i \pi  h_0 h_1 (2 v-1)\nonumber\\
&-\frac{1}{2} i \pi  h_0^2 h_1 (2 v+1)-\frac{1}{6} i \pi  h_1^3 (2 v+9)-4 i \pi  h_3 (2 v+5)
+\frac{11}{12} \pi ^2 h_0^2 (4 v-3)+8 h_0 h_3 (4 v-5)\nonumber\\
&+\frac{1}{8} h_0^4 (4 v+1)+\frac{1}{2} i \pi  h_0 h_1^2 (6 v-1)-\frac{2}{3} i \pi ^3 (4 v+1)+\frac{1}{6} i \pi ^3 h_0 (2 v+17)-16 h_3 (v-2)\nonumber\\
&+h_0 h_1 h_2 (16 v-8)+\frac{1}{6} \pi ^2 h_0 h_1 (44 v-31)+h_2^2 (14 v-7)+16 i \pi  h_2 (2 v-1)\nonumber\\
&+h_0 (h_{21} (20-40 v)+64 v)+4 h_1 (4 h_{21} v+h_{21}+16 (v-1))-4 i \pi  (h_{21} (2 v-7)-16)\nonumber\\
&+h_0 (-30 v-3) \zeta_{3}+h_1 (21-30 v) \zeta_{3}-2 (8 h_{21} (v+1)+6 h_{211} (4 v+1)+7 h_{22} (2 v-1))\nonumber\\
&+16 v \zeta_{3}+i \pi  (8 v-61) \zeta_{3}\,.
\end{align}
This result is consistent with Ref.~\cite{Kudashkin:2017skd}
after a proper analytic continuation.

One remarkable feature of the result is
that it contains terms proportionals to $1/m_t$.
Higher $m_t$-order correction also contain odd-power terms.
These odd-power terms come from Region~2~and~3:
\begin{align}
&\lim_{\epsilon,\delta_7,\delta_6,\delta_5,\delta_4,\delta_3,\delta_2,\delta_1\to0}
\left(
\mathcal{T}^{(2)}
+\mathcal{T}^{(3)}
\right)
\times e^{2\epsilon\gamma_E}
\nonumber\\
&\qquad =
\frac{i\pi^3}{m_ts^{\frac{5}{2}}\sqrt{v}\sqrt{1-v}}
\left( \frac{1}{\epsilon}-2\log (m_t^2)-10\log 2\right)
+m_t^0~\Delta (1/\delta_j)
+\mathcal{O}(\epsilon,m_t)
\,,
\label{two-tot1}
\end{align}
where $\Delta (1/\delta_j)$
has poles in $\delta_j$,
and these poles are cancelled by the contributions from the other regions.

%%%%%%%%%%%%%%%%%%%%%%%%%%%%%%%%%%%%%%%%%%%%%%%%%%%%%%%
\subsection{Other Master Integrals}
\label{ss:dots}

\subsubsection*{Seven-Line Integrals}

After the IBP-reduction described in Ref.~\cite{Davies:2018qvx},
we find that there are four more diagrams
which have seven internal lines.
We consider 
$J^\mathrm{NPL1}_{2,1,1,1,1,1,1}$,
$J^\mathrm{NPL1}_{1,1,2,1,1,1,1}$,
$J^\mathrm{NPL1}_{1,1,1,2,1,1,1}$,
$J^\mathrm{NPL1}_{1,1,1,1,1,2,1}$
which are used to calculate 
the Higgs pair production cross section.
(For the detail, see Ref.~\cite{Davies:2018qvx}.)
These integrals can be expressed by
the proper shifts of $\delta_j$:
$\delta_1\to\delta_1+1$ for $J^\mathrm{NPL1}_{2,1,1,1,1,1,1}$,
for example.

\subsubsection*{Six-Line Integrals}

We can use the method described in Subsection~\ref{ss:fewer}
to compute the integral with fewer lines.
For example,
$J^\mathrm{NPL1}_{1,1,1,1,1,1,0}$
is obtained 
by shifting $\delta_7\to\delta_7-1$
and repeat the same procedure described above.
Note that the analytic continuation of $\delta_j$ may change
due to the shift.
For example, 
the template integral of Region 12~\eqref{template-npl12}
can be regularized with $\delta_{i>0}=0$,
whereas
$J^\mathrm{NPL1}_{1,1,1,1,1,1,0}$
requires a non-zero $\delta_7$ to regularize that template integral.

%%%%%%%%%%%%%%%%%%%%%%%%%%%%%%%%%%%%%%%%%%%%%%%%%%%%%%%
\section{Summary}
\label{ss:sum}

Asymptotic expansion
is useful to extract information from multi-scale Feynman integrals,
which are difficult to solve exactly.
The method of regions 
plays an essential role 
in this extraction.
The crucial part of the method of regions 
is to reveal the relevant regions correctly,
and a naive application of the conventional method
fails in the case of non-planar integrals
for which the second Symanzik polynomial
does not have a definite sign.
We solve this problem by performing an analytic continuation of the Mandelstam variables 
such that the second Symanzik polynomial is positive definite.

We show the applicability of the method of regions
by the explicit calculation of the master integrals 
of the Higgs pair production cross section at two-loop order,
in the high energy limit.
It is straightforward to extend our calculation to other 
four-point two-loop integral which satisfy $q_i^2\ll m^2\ll S,T,U$
where $q_i$ are the external momenta and $m$ is the mass of the internal lines.
We anticipate that our idea to make the second Symanzik polynomial positive definite 
works in more general cases.

In addition to solving the issue of the sign of the second Symanzik polynomial,
we formulate the procedure of the calculation in a systematic way.
The contribution from each region
is expressed in terms of Mellin-Barnes integrals,
and a way to solve them is presented.
The procedure presented here 
to solve the Mellin-Barnes integrals
beyond the Barnes lemmas is not applicable to the general case,
although it is sufficient to solve our master integrals completely.
The automatization of this part is
a future project.

As a by-product of introducing the positive Mandelstam variables,
it becomes easier to obtain the crossed integrals,
since the crossing of the positive Mandelstam variables 
does not cross any branch cut.

We compute the first few terms of the series in $m_t$,
and the higher order terms can be obtained by the use of the $m_t$-differential equations.
In this sense, our results can be considered as the boundary conditions
of the differential equations with respect to $m_t$.

\section*{Acknowledgements}

The author is grateful to
Joshua Davies, Matthias Steinhauser, and David Wellmann
for valuable collaborative work,
and Hjalte Frellesvig, Christopher Wever
for useful discussions.
The author sincerely thanks 
Matthias Steinhauser, Joshua Davies, Vladimir Smirnov,
and Yukinari Sumino
for carefully reading the manuscript and a lot of useful comments.

\appendix

%%%%%%%%%%%%%%%%%%%%%%%%%%%%%%%%%%%%%%%%%%%%%%%%%%%%%%%
\section{Mellin-Barnes Integrals}
\label{ss:introMB}

\begin{figure}
\centering
% \begin{tabular}{cc}
% \begin{minipage}{0.45\hsize}
% \centering
%\begin{tikzpicture}[baseline={(0,0)},scale=0.2]
%\draw (-12,8) node {(a)};
%\draw [->] (1,-10)--(1,-7);
%\draw (0,-6) arc [start angle=90, end angle=0, radius=1];
%\draw (-6,0) arc [start angle=180, end angle=270, radius=6];
%\draw (-5,1) arc [start angle=90, end angle=180, radius=1];
%\draw (0,0) sin (-1,1) cos (-2,0) sin (-3,-1) cos (-4,0) sin (-5,1);
%\draw (0,0) sin (1,-1) cos (2,0) sin (3,1) cos (4,0) sin (5,-1);
%\draw (5,-1) arc [start angle=270, end angle=360, radius=1];
%\draw (6,0) arc [start angle=0, end angle=90, radius=6];
%\draw (0,6) arc [start angle=270, end angle=180, radius=1];
%\draw [->] (-1,7)--(-1,8);
%\draw [-] (-1,8)--(-1,10);
%\draw (1,0) node {$\times$};
%\draw (5,0) node {$\times$};
%\draw (-3,0) node {$\times$};
%\draw (-7,0) node {$\times$};
%\draw (-11,0) node {$\times$};
%\draw (-15,0) node {$\times$};
%\draw (-5,0) node {$\circ$};
%\draw (-1,0) node {$\circ$};
%\draw (3,0) node {$\circ$};
%\draw (7,0) node {$\circ$};
%\draw (11,0) node {$\circ$};
%\draw (15,0) node {$\circ$};
%\end{tikzpicture}
% \end{minipage}
% \begin{minipage}{0.45\hsize}
% \centering
%\begin{tikzpicture}[baseline={(0,0)},scale=0.2]
%\draw (-12,8) node {(b)};
%\draw [->] (0,-10)--(0,-7);
%\draw [->] (0,-7)--(0,8);
%\draw [-] (0,8)--(0,10);
%\draw (-3,0) node {$\times$};
%\draw (-7,0) node {$\times$};
%\draw (-11,0) node {$\times$};
%\draw (-15,0) node {$\times$};
%\draw (3,0) node {$\circ$};
%\draw (7,0) node {$\circ$};
%\draw (11,0) node {$\circ$};
%\draw (15,0) node {$\circ$};
%\end{tikzpicture}
% \end{minipage}
% \end{tabular}
\includegraphics[width=0.8\textwidth]{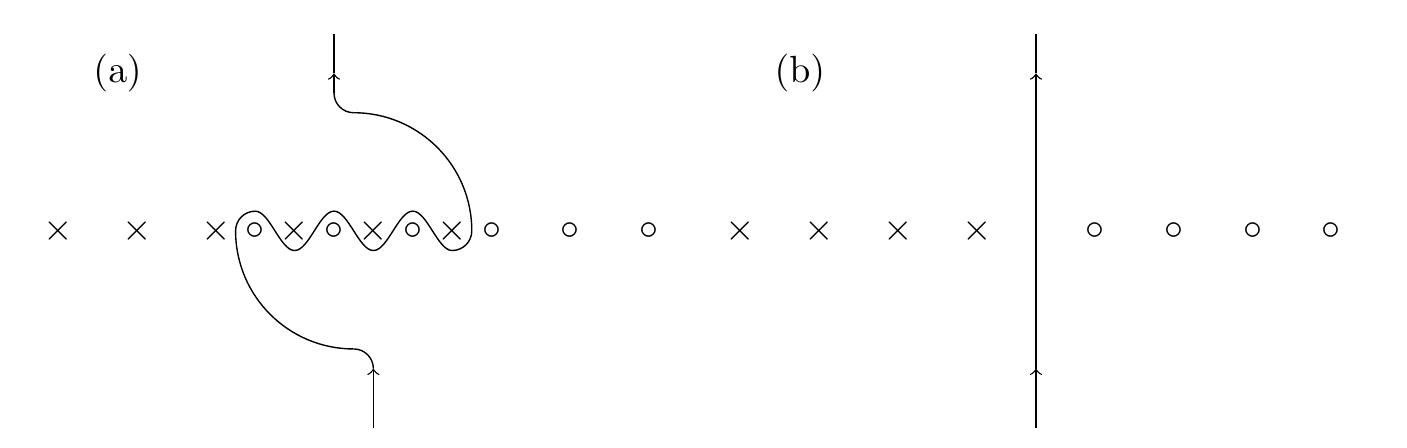}
\caption{
A circle ($\circ$) represents the right pole
and a cross ($\times$) represents the left pole.
(a) An example of the contour in Eq.~\eqref{mb1}
with $\lambda =-5/2$.
(b) An example of the contour in Eq.~\eqref{mb2}
with $\lambda =2, c_0=1$.
}
\label{fig:contour1}
\end{figure}

Here we explain elementary aspects 
of the Mellin-Barnes integrals.
The basic tool is the identity 
\begin{align}
\frac{1}{\left( A+B \right)^{\lambda}}
=
\int_C \frac{\mathrm{d}z}{2\pi i}~
\frac{B^z}{A^{\lambda+z}}~
\frac{\Gamma [-z,\lambda+z]}{\Gamma (\lambda)}
\label{mb1}
\,,
\end{align}
where the integration contour $C$
satisfies the following three properties:
\renewcommand{\labelitemi}{$\circ$}
\begin{itemize}
\item $C$ runs from $-i\infty$ to $+i\infty$.
\item The poles of $\Gamma (-z)$ lie on the right side of $C$.
(We call them right poles.)
\item The poles of $\Gamma (\lambda +z)$ lie on the left side of $C$.
(We call them left poles.)
\end{itemize}
The separation of the left poles and right poles is possible
if $\lambda \not \in \{0,-1,-2,-3,...\}$.
For example, when $\lambda=-7/2$,
some of the left poles lie on the right side of the left-most right pole,
but there exists a contour which separates the left poles and the right poles 
in the proper manner. [See Fig.~\ref{fig:contour1}~(a).]
In our calculation,
$\lambda$ is a combination of
$\epsilon$ and $\delta_j$ and integration variables
of other Mellin-Barnes integrals,
and $A,B$ are combinations
of the positive Mandelstam variables
and the alpha parameters.
In particular,
$A>0$ and $B>0$ are always satisfied.
We choose $\delta_j$ to
ensure the condition $\lambda \not \in \{0,-1,-2,-3,...\}$.\footnote{
If some of the left poles and right poles merge
for any choice of $\delta_j$,
it is necessary to compensate
the contributions of the merged poles
by adding or subtracting the residue of the poles.
However, 
it turns out that 
the merger of poles does not happen 
in our cases.
}
When $\mathrm{Re}(\lambda )>0$,
it is possible to set $C$ as a straight line along the imaginary axis
[See Fig.~\ref{fig:contour1}~(b).]
\begin{align}
\frac{1}{\left( A+B \right)^{\lambda}}
=
\int_{-c_0 -i\infty}^{-c_0+i\infty}
 \frac{\mathrm{d}z}{2\pi i}~
\frac{B^z}{A^{\lambda+z}}~
\frac{\Gamma [-z,\lambda+z]}{\Gamma (\lambda)}
\qquad 0<c_0<\mathrm{Re}(\lambda)
\label{mb2}
\,.
\end{align}
The integrand converges rapidly to zero at $\mathrm{Im}(z)=\pm\infty$
rapidly so that we can shift the 
end points of the integration.

It is sometimes difficult to fix the contour $C$ immediately 
when the Mellin-Barnes representation is introduced
because additional convergence conditions
concerning the $\alpha_i$-integration
should be taken into account.
For example,
the relations 
\begin{align}
&\int _0^\infty \mathrm{d}\alpha~\alpha ^a
e^{-A\alpha }
\myeq
\Gamma(1+a)A^{-1-a}
\,,
\label{form1}
\\
&\int _0^\infty \mathrm{d}\alpha~\alpha ^a(A+B\alpha )^b
\myeq
A^{1+a+b}B^{-1-a}
\frac{\Gamma [1+a,-1-a-b]}{\Gamma (-b)}
\,,
\label{form2}
\end{align}
require $\mathrm{Re}(a)>-1$ and 
$\mathrm{Re}(a+b)<-1$,
and if $a$ or $b$ contains the Mellin-Barnes integration variable $z$,
$C$ should be chosen accordingly.
This kind of convergence condition appears every time an integration is performed,
and it can happen that an analytic continuation of the regularization parameters
is necessary between two successive integrations.
Therefore it is necessary to deform $C$ smoothly 
in each step of the integration
assuming a flexible contour like Fig.~\ref{fig:contour1}~(a)
in order to ensure convergence.
Note that after the integration, the analytic continuation of $\Gamma$-functions
in the left hand sides of Eqs.~\eqref{form1},~\eqref{form2} is possible.

After the integration with respect to all $\alpha_i$ is solved,
we deform $C$ to a straight line 
by tuning $\delta_j$ to separate the right poles and the left poles 
as in Fig.~\ref{fig:contour1}~(b).
We obtain an expression which is valid for certain values of $\delta_j$.
In order to take the limit $\delta_j\to 0$,
it is necessary to apply a final analytic continuation of $\delta_j$ 
from their initial values
into a neighborhood around 0.
This time, we allow poles to cross $C$.
For this procedure, 
we use the package \texttt{MB.m}~\cite{Czakon:2005rk}.

%%%%%%%%%%%%%%%%%%%%%%%%%%%%%%%%%%%%%%%%%%%%%%%%%%%%%%
\section{Template Integrals for the Two-Loop Master Integrals of Higgs Pair Production}
\label{app:temp}

The results for the template integrals listed here
can be found in the ancillary file to this paper~\cite{anci},
in a computer-readable format.

\subsection*{Template Integrals of NPL2}

The template integrals of $I^\mathrm{NPL2}$
are
\begin{align}
\mathcal{T}^{(2)}
=&
\frac{1}{2}
(m_t^2)^{-(1+4\epsilon+\delta_{112344556})/2}
S^{-(3+\delta_{23\bar 677})/2}
T^{-(1+\delta_{\bar 2 36})/2}
U^{-(1+\delta_{2\bar 36})/2}
\nonumber\\
&\qquad \times
\frac{\Gamma [
\delta_{\bar0\bar5\bar6},
\frac{1+\delta_{23\bar 6}}{2},
\frac{1+\delta_{2\bar 36}}{2},
\frac{1+\delta_{\bar 236}}{2},
\frac{1+\delta_{\bar 2\bar 3556}}{2},
\frac{1+\delta_{00112\bar 3556}}{2},
\frac{1+\delta_{00\bar 2344556}}{2},
\frac{1+\delta_{0000112344556}}{2}
]}
{\Gamma [
\delta_1+1,
\delta_2+1,
\delta_3+1,
\delta_4+1,
\delta_5+1,
\delta_6+1,
\delta_{0014556}+1,
\frac{1-\delta_{00236}}{2}
]}
\label{template-npl2-2}
\\
\mathcal{T}^{(3)}
=&
\frac{1}{2}
(m_t^2)^{-(1+4\epsilon+\delta_{122334677})/2}
S^{-(3+\delta_{1455\bar 6})/2}
T^{-(1+\delta_{1\bar 46})/2}
U^{-(1+\delta_{\bar 146})/2}
\nonumber\\
&\qquad \times
\frac{\Gamma [
\delta_{\bar0\bar6\bar7},
\frac{1+\delta_{14\bar 6}}{2},
\frac{1+\delta_{1\bar 46}}{2},
\frac{1+\delta_{\bar 146}}{2},
\frac{1+\delta_{\bar 1\bar 4677}}{2},
\frac{1+\delta_{00122\bar 4677}}{2},
\frac{1+\delta_{00\bar 1334677}}{2},
\frac{1+\delta_{0000122334677}}{2}
]}
{\Gamma [
\delta_1+1,
\delta_2+1,
\delta_3+1,
\delta_4+1,
\delta_6+1,
\delta_7+1,
\delta_{0023677}+1,
\frac{1-\delta_{00146}}{2}
]}
\label{template-npl2-3}
\\
\mathcal{T}^{(4)}=&
\int\mathrm{d}z_1
\frac{
(m_t^2)^{-\delta_{001234}}
\Gamma[\delta_{012},\delta_{034},z_{\bar{1}}+\delta_{3\bar{5}\bar{6}}-1,z_{\bar{1}}+\delta_{1\bar{6}\bar{7}}-1,z_{\bar{1}},z_{1}+\delta_{2\bar{5}},z_{1}+\delta_{6}+1,z_{1}+\delta_{4\bar{7}}]}
{S^{2+\delta_{57}}T^{1+z_{1}+\delta_{6}}U^{z_{\bar{1}}}
\Gamma[\delta_{1}+1,\delta_{2}+1,\delta_{3}+1,\delta_{4}+1,\delta_{6}+1,\delta_{12\bar{5}\bar{6}\bar{7}}-1,\delta_{34\bar{5}\bar{6}\bar{7}}-1]}\label{template-npl2-4}
\\
\mathcal{T}^{(5)}=&
\int\mathrm{d}z_1
\frac{\Gamma[\delta_{037},\delta_{001237},\delta_{\bar{4}7},z_{\bar{1}}+\delta_{0267}+1,z_{\bar{1}},z_{1}+\delta_{1\bar{4}\bar{6}}-1,z_{1}+\delta_{\bar{0}\bar{4}\bar{5}\bar{6}}-1,z_{1}+\delta_{\bar{0}\bar{6}\bar{7}}]}
{(m_t^2)^{\delta_{001237}}S^{2+\delta_{45}}T^{1+\delta_{6}}
\Gamma[\delta_{1}+1,\delta_{2}+1,\delta_{3}+1,\delta_{\bar{0}\bar{4}\bar{5}\bar{6}}-1,\delta_{7}+1,\delta_{00123\bar{4}77},z_{1}+\delta_{\bar{0}\bar{4}\bar{6}}]}\label{template-npl2-5}
\\
\mathcal{T}^{(6)}=&
\int\mathrm{d}z_1
\frac{\Gamma[\delta_{\bar{3}5},\delta_{045},\delta_{001245},z_{\bar{1}}+\delta_{0156}+1,z_{\bar{1}},z_{1}+\delta_{2\bar{3}\bar{6}}-1,z_{1}+\delta_{\bar{0}\bar{5}\bar{6}},z_{1}+\delta_{\bar{0}\bar{3}\bar{6}\bar{7}}-1]}
{(m_t^2)^{\delta_{001245}}S^{2+\delta_{37}}U^{1+\delta_{6}}
\Gamma[\delta_{1}+1,\delta_{2}+1,\delta_{4}+1,\delta_{5}+1,\delta_{0012\bar{3}455},\delta_{\bar{0}\bar{3}\bar{6}\bar{7}}-1,z_{1}+\delta_{\bar{0}\bar{3}\bar{6}}]}\label{template-npl2-6}
\\
\mathcal{T}^{(7)}=&
\int\mathrm{d}z_1
\frac{\Gamma[\delta_{015},\delta_{\bar{2}5},\delta_{001345},z_{\bar{1}}+\delta_{0456}+1,z_{\bar{1}},z_{1}+\delta_{\bar{2}3\bar{6}}-1,z_{1}+\delta_{\bar{0}\bar{5}\bar{6}},z_{1}+\delta_{\bar{0}\bar{2}\bar{6}\bar{7}}-1]}
{(m_t^2)^{\delta_{001345}}S^{2+\delta_{27}}T^{1+\delta_{6}}
\Gamma[\delta_{1}+1,\delta_{3}+1,\delta_{4}+1,\delta_{5}+1,\delta_{001\bar{2}3455},\delta_{\bar{0}\bar{2}\bar{6}\bar{7}}-1,z_{1}+\delta_{\bar{0}\bar{2}\bar{6}}]}\label{template-npl2-7}
\\
\mathcal{T}^{(8)}=&
\frac{\Gamma[\delta_{\bar{0}\bar{5}\bar{6}},\delta_{\bar{2}\bar{3}6}-1,\delta_{01\bar{3}56},\delta_{001456},\delta_{0\bar{2}456},\delta_{\bar{0}\bar{6}\bar{7}}]}
{(m_t^2)^{\delta_{001456}}S^{1+\delta_{7}}T^{1+\delta_{3}}U^{1+\delta_{2}}
\Gamma[\delta_{1}+1,\delta_{4}+1,\delta_{\bar{0}\bar{6}}+1,\delta_{6}+1,\delta_{001\bar{2}\bar{3}45566},\delta_{\bar{0}\bar{2}\bar{3}\bar{7}}-1]}\label{template-npl2-8}
\\
\mathcal{T}^{(9)}=&
\int\mathrm{d}z_1
\frac{\Gamma[\delta_{\bar{1}7},\delta_{027},\delta_{002347},z_{\bar{1}}+\delta_{0367}+1,z_{\bar{1}},z_{1}+\delta_{\bar{1}4\bar{6}}-1,z_{1}+\delta_{\bar{0}\bar{1}\bar{5}\bar{6}}-1,z_{1}+\delta_{\bar{0}\bar{6}\bar{7}}]}
{(m_t^2)^{\delta_{002347}}S^{2+\delta_{15}}U^{1+\delta_{6}}
\Gamma[\delta_{2}+1,\delta_{3}+1,\delta_{4}+1,\delta_{\bar{0}\bar{1}\bar{5}\bar{6}}-1,\delta_{7}+1,\delta_{00\bar{1}23477},z_{1}+\delta_{\bar{0}\bar{1}\bar{6}}]}\label{template-npl2-9}
\\
\mathcal{T}^{(10)}=&
\frac{\Gamma[\delta_{\bar{0}\bar{5}\bar{6}},\delta_{\bar{1}\bar{4}6}-1,\delta_{\bar{0}\bar{6}\bar{7}},\delta_{0\bar{1}367},\delta_{002367},\delta_{02\bar{4}67}]}
{(m_t^2)^{\delta_{002367}}S^{1+\delta_{5}}T^{1+\delta_{1}}U^{1+\delta_{4}}
\Gamma[\delta_{2}+1,\delta_{3}+1,\delta_{\bar{0}\bar{1}\bar{4}\bar{5}}-1,\delta_{\bar{0}\bar{6}}+1,\delta_{6}+1,\delta_{00\bar{1}23\bar{4}6677}]}\label{template-npl2-10}
\\
\mathcal{T}^{(11)}=&
\int\mathrm{d}z_1\mathrm{d}z_2\mathrm{d}z_3\mathrm{d}z_4
\frac{\Gamma[\delta_{012},z_{\bar{1}},z_{\bar{2}},z_{12}+\delta_{6}+1,z_{\bar{3}},z_{13}+\delta_{2}+1,z_{23}+\delta_{3}+1,z_{\bar{1}\bar{2}\bar{3}}+\delta_{\bar{0}\bar{3}\bar{4}\bar{5}\bar{6}}-2]}
{(m_t^2)^{\delta_{012}}S^{3+z_{12}+\delta_{034567}}T^{z_{\bar{2}}}U^{z_{\bar{1}}}
\Gamma[\delta_{1}+1,\delta_{2}+1,\delta_{3}+1,\delta_{4}+1]}\nonumber\\
&\qquad\times
\frac{\Gamma[z_{\bar{1}\bar{3}\bar{4}}+\delta_{\bar{0}1\bar{3}\bar{4}\bar{5}\bar{6}\bar{7}}-2,z_{\bar{4}},z_{34}+\delta_{5}+1,z_{1234}+\delta_{034567}+3,z_{\bar{2}\bar{3}\bar{4}}+\delta_{\bar{0}\bar{3}\bar{5}\bar{6}\bar{7}}-2]}
{\Gamma[\delta_{5}+1,\delta_{6}+1,\delta_{\bar{0}\bar{0}\bar{3}\bar{4}\bar{5}\bar{6}\bar{7}}-1,\delta_{7}+1,z_{\bar{4}}+\delta_{\bar{0}12\bar{3}\bar{4}\bar{5}\bar{6}\bar{7}}-1]}\label{template-npl2-11}
\\
\mathcal{T}^{(12)}=&
\int\mathrm{d}z_1\mathrm{d}z_2\mathrm{d}z_3\mathrm{d}z_4
\frac{\Gamma[\delta_{034},z_{\bar{1}},z_{\bar{2}},z_{12}+\delta_{6}+1,z_{\bar{3}},z_{13}+\delta_{3}+1,z_{23}+\delta_{2}+1,z_{\bar{1}\bar{2}\bar{3}}+\delta_{\bar{0}\bar{1}\bar{2}\bar{5}\bar{6}}-2]}
{(m_t^2)^{\delta_{034}}S^{3+z_{12}+\delta_{012567}}T^{z_{\bar{1}}}U^{z_{\bar{2}}}
\Gamma[\delta_{1}+1,\delta_{2}+1,\delta_{3}+1,\delta_{4}+1]}\nonumber\\
&\qquad\times
\frac{\Gamma[z_{\bar{1}\bar{3}\bar{4}}+\delta_{\bar{0}\bar{1}\bar{2}4\bar{5}\bar{6}\bar{7}}-2,z_{\bar{4}},z_{34}+\delta_{5}+1,z_{1234}+\delta_{012567}+3,z_{\bar{2}\bar{3}\bar{4}}+\delta_{\bar{0}\bar{2}\bar{5}\bar{6}\bar{7}}-2]}
{\Gamma[\delta_{5}+1,\delta_{6}+1,\delta_{\bar{0}\bar{0}\bar{1}\bar{2}\bar{5}\bar{6}\bar{7}}-1,\delta_{7}+1,z_{\bar{4}}+\delta_{\bar{0}\bar{1}\bar{2}34\bar{5}\bar{6}\bar{7}}-1]}
\,.
\label{template-npl2-12}
\end{align}

\subsection*{Template Integrals of PL1}

The template integrals of 
 $I^\mathrm{PL1}=J^\mathrm{PL1}_{1,1,1,1,1,1,1}$
defined in Eq.~\eqref{pl-1}
are
\begin{align}
\mathcal{T}^{(2)}=&
\int\mathrm{d}z_1
\frac{\Gamma[\delta_{012},\delta_{045},z_{\bar{1}}+\delta_{1\bar{3}\bar{7}}-1,z_{\bar{1}}+\delta_{4\bar{6}\bar{7}}-1,z_{\bar{1}},z_{1}+\delta_{2}+1,z_{1}+\delta_{5}+1,z_{1}+\delta_{7}+1]}
{(m_t^2)^{\delta_{001245}}S^{3+z_{1}+\delta_{367}}T^{z_{\bar{1}}}
\Gamma[\delta_{1}+1,\delta_{2}+1,\delta_{4}+1,\delta_{5}+1,\delta_{12\bar{3}\bar{7}},\delta_{45\bar{6}\bar{7}},\delta_{7}+1]}\label{template-pl1-2}\\
\mathcal{T}^{(3)}=&
\frac{\Gamma[\delta_{012},\delta_{1\bar{3}},\delta_{\bar{4}6},\delta_{056},\delta_{2\bar{7}},\delta_{5\bar{7}}]}
{(m_t^2)^{\delta_{001256}}S^{2+\delta_{34}}T^{1+\delta_{7}}
\Gamma[\delta_{1}+1,\delta_{2}+1,\delta_{5}+1,\delta_{6}+1,\delta_{12\bar{3}\bar{7}},\delta_{\bar{4}56\bar{7}}]}\label{template-pl1-3}\\
\mathcal{T}^{(4)}=&
\int\mathrm{d}z_1\mathrm{d}z_2
\frac{\Gamma[\delta_{067},z_{\bar{1}},z_{1}+\delta_{\bar{0}\bar{3}\bar{4}\bar{5}}-1,z_{1}+\delta_{\bar{5}7},z_{\bar{2}}+\delta_{017},z_{\bar{1}\bar{2}}+\delta_{001567},z_{\bar{2}},z_{2}+\delta_{\bar{0}\bar{7}}+1]}
{(m_t^2)^{\delta_{001267}}S^{2+\delta_{34}}T^{1+\delta_{5}}
\Gamma[\delta_{1}+1,\delta_{2}+1,\delta_{\bar{0}\bar{3}\bar{4}\bar{5}}-1,\delta_{6}+1,\delta_{7}+1,z_{1}+\delta_{\bar{0}\bar{3}\bar{5}}]}\nonumber\\
&\times\frac{\Gamma[z_{12}+\delta_{2\bar{5}},z_{12}+\delta_{\bar{0}\bar{3}\bar{5}}]}
{\Gamma[z_{\bar{2}}+\delta_{001677},z_{12}+\delta_{\bar{0}\bar{5}}+1]}
\label{template-pl1-4}\\
\mathcal{T}^{(5)}=&
\int\mathrm{d}z_1\mathrm{d}z_2
\frac{\Gamma[\delta_{017},z_{\bar{1}},z_{1}+\delta_{\bar{0}\bar{2}\bar{3}\bar{4}}-1,z_{1}+\delta_{\bar{2}7},z_{\bar{2}}+\delta_{067},z_{\bar{1}\bar{2}}+\delta_{001267},z_{\bar{2}},z_{2}+\delta_{\bar{0}\bar{7}}+1]}
{(m_t^2)^{\delta_{001567}}S^{2+\delta_{34}}T^{1+\delta_{2}}
\Gamma[\delta_{1}+1,\delta_{\bar{0}\bar{2}\bar{3}\bar{4}}-1,\delta_{5}+1,\delta_{6}+1,\delta_{7}+1,z_{1}+\delta_{\bar{0}\bar{2}\bar{4}}]}\nonumber\\
&\times\frac{\Gamma[z_{12}+\delta_{\bar{0}\bar{2}\bar{4}},z_{12}+\delta_{\bar{2}5}]}
{\Gamma[z_{\bar{2}}+\delta_{001677},z_{12}+\delta_{\bar{0}\bar{2}}+1]}
\label{template-pl1-5}\\
\mathcal{T}^{(6)}=&
\frac{\Gamma[\delta_{\bar{1}3},\delta_{023},\delta_{045},\delta_{4\bar{6}},\delta_{2\bar{7}},\delta_{5\bar{7}}]}
{(m_t^2)^{\delta_{002345}}S^{2+\delta_{16}}T^{1+\delta_{7}}
\Gamma[\delta_{2}+1,\delta_{3}+1,\delta_{4}+1,\delta_{5}+1,\delta_{\bar{1}23\bar{7}},\delta_{45\bar{6}\bar{7}}]}\label{template-pl1-6}\\
\mathcal{T}^{(7)}=&
\int\mathrm{d}z_1\mathrm{d}z_2
\frac{\Gamma[\delta_{047},z_{\bar{1}},z_{1}+\delta_{\bar{0}\bar{1}\bar{5}\bar{6}}-1,z_{1}+\delta_{\bar{5}7},z_{\bar{2}}+\delta_{037},z_{\bar{1}\bar{2}}+\delta_{003457},z_{\bar{2}},z_{2}+\delta_{\bar{0}\bar{7}}+1]}
{(m_t^2)^{\delta_{002347}}S^{2+\delta_{16}}T^{1+\delta_{5}}
\Gamma[\delta_{2}+1,\delta_{3}+1,\delta_{4}+1,\delta_{\bar{0}\bar{1}\bar{5}\bar{6}}-1,\delta_{7}+1,z_{1}+\delta_{\bar{0}\bar{1}\bar{5}}]}\nonumber\\
&\times\frac{\Gamma[z_{12}+\delta_{\bar{0}\bar{1}\bar{5}},z_{12}+\delta_{2\bar{5}}]}
{\Gamma[z_{\bar{2}}+\delta_{003477},z_{12}+\delta_{\bar{0}\bar{5}}+1]}
\label{template-pl1-7}\\
\mathcal{T}^{(8)}=&
\int\mathrm{d}z_1
\frac{\Gamma[\delta_{023},\delta_{056},z_{\bar{1}}+\delta_{\bar{1}3\bar{7}}-1,z_{\bar{1}}+\delta_{\bar{4}6\bar{7}}-1,z_{\bar{1}},z_{1}+\delta_{2}+1,z_{1}+\delta_{5}+1,z_{1}+\delta_{7}+1]}
{(m_t^2)^{\delta_{002356}}S^{3+z_{1}+\delta_{147}}T^{z_{\bar{1}}}
\Gamma[\delta_{2}+1,\delta_{3}+1,\delta_{5}+1,\delta_{6}+1,\delta_{\bar{1}23\bar{7}},\delta_{\bar{4}56\bar{7}},\delta_{7}+1]}\label{template-pl1-8}\\
\mathcal{T}^{(9)}=&
\int\mathrm{d}z_1\mathrm{d}z_2
\frac{\Gamma[\delta_{037},z_{\bar{1}},z_{1}+\delta_{\bar{0}\bar{1}\bar{2}\bar{6}}-1,z_{1}+\delta_{\bar{2}7},z_{\bar{2}}+\delta_{047},z_{\bar{1}\bar{2}}+\delta_{002347},z_{\bar{2}},z_{2}+\delta_{\bar{0}\bar{7}}+1]}
{(m_t^2)^{\delta_{003457}}S^{2+\delta_{16}}T^{1+\delta_{2}}
\Gamma[\delta_{3}+1,\delta_{4}+1,\delta_{5}+1,\delta_{\bar{0}\bar{1}\bar{2}\bar{6}}-1,\delta_{7}+1,z_{1}+\delta_{\bar{0}\bar{2}\bar{6}}]}\nonumber\\
&\times\frac{\Gamma[z_{12}+\delta_{\bar{2}5},z_{12}+\delta_{\bar{0}\bar{2}\bar{6}}]}
{\Gamma[z_{\bar{2}}+\delta_{003477},z_{12}+\delta_{\bar{0}\bar{2}}+1]}
\label{template-pl1-9}\\
\mathcal{T}^{(10)}=&
\int\mathrm{d}z_1\mathrm{d}z_2
\frac{\Gamma[\delta_{012},z_{\bar{1}},z_{1}+\delta_{2}+1,z_{1}+\delta_{5}+1,z_{\bar{1}}+\delta_{\bar{0}\bar{5}\bar{6}\bar{7}}-1,z_{\bar{2}},z_{2}+\delta_{1\bar{3}},z_{12}+\delta_{7}+1]}
{(m_t^2)^{\delta_{012}}S^{3+z_{1}+\delta_{034567}}T^{z_{\bar{1}}}
\Gamma[\delta_{1}+1,\delta_{2}+1,\delta_{4}+1,\delta_{5}+1,\delta_{6}+1,\delta_{\bar{0}\bar{0}\bar{4}\bar{5}\bar{6}\bar{7}}]}\nonumber\\
&\times\frac{\Gamma[z_{12}+\delta_{04567}+2,z_{\bar{1}\bar{2}}+\delta_{\bar{0}\bar{4}\bar{5}\bar{7}}-1]}
{\Gamma[\delta_{7}+1,z_{12}+\delta_{12\bar{3}}+1]}
\label{template-pl1-10}\\
\mathcal{T}^{(11)}=&
\int\mathrm{d}z_1\mathrm{d}z_2
\frac{\Gamma[\delta_{023},z_{\bar{1}},z_{1}+\delta_{2}+1,z_{1}+\delta_{5}+1,z_{\bar{1}}+\delta_{\bar{0}\bar{4}\bar{5}\bar{7}}-1,z_{\bar{2}},z_{2}+\delta_{\bar{1}3},z_{12}+\delta_{7}+1]}
{(m_t^2)^{\delta_{023}}S^{3+z_{1}+\delta_{014567}}T^{z_{\bar{1}}}
\Gamma[\delta_{2}+1,\delta_{3}+1,\delta_{4}+1,\delta_{5}+1,\delta_{6}+1,\delta_{\bar{0}\bar{0}\bar{4}\bar{5}\bar{6}\bar{7}}]}\nonumber\\
&\times\frac{\Gamma[z_{12}+\delta_{04567}+2,z_{\bar{1}\bar{2}}+\delta_{\bar{0}\bar{5}\bar{6}\bar{7}}-1]}
{\Gamma[\delta_{7}+1,z_{12}+\delta_{\bar{1}23}+1]}
\label{template-pl1-11}\\
\mathcal{T}^{(12)}=&
\int\mathrm{d}z_1\mathrm{d}z_2
\frac{\Gamma[\delta_{045},z_{\bar{1}},z_{1}+\delta_{2}+1,z_{1}+\delta_{5}+1,z_{\bar{1}}+\delta_{\bar{0}\bar{2}\bar{3}\bar{7}}-1,z_{\bar{2}},z_{2}+\delta_{4\bar{6}},z_{12}+\delta_{7}+1]}
{(m_t^2)^{\delta_{045}}S^{3+z_{1}+\delta_{012367}}T^{z_{\bar{1}}}
\Gamma[\delta_{1}+1,\delta_{2}+1,\delta_{3}+1,\delta_{4}+1,\delta_{5}+1,\delta_{\bar{0}\bar{0}\bar{1}\bar{2}\bar{3}\bar{7}}]}\nonumber\\
&\times\frac{\Gamma[z_{12}+\delta_{01237}+2,z_{\bar{1}\bar{2}}+\delta_{\bar{0}\bar{1}\bar{2}\bar{7}}-1]}
{\Gamma[\delta_{7}+1,z_{12}+\delta_{45\bar{6}}+1]}
\label{template-pl1-12}\\
\mathcal{T}^{(13)}=&
\int\mathrm{d}z_1\mathrm{d}z_2
\frac{\Gamma[\delta_{056},z_{\bar{1}},z_{1}+\delta_{2}+1,z_{1}+\delta_{5}+1,z_{\bar{1}}+\delta_{\bar{0}\bar{1}\bar{2}\bar{7}}-1,z_{\bar{2}},z_{2}+\delta_{\bar{4}6},z_{12}+\delta_{7}+1]}
{(m_t^2)^{\delta_{056}}S^{3+z_{1}+\delta_{012347}}T^{z_{\bar{1}}}
\Gamma[\delta_{1}+1,\delta_{2}+1,\delta_{3}+1,\delta_{5}+1,\delta_{6}+1,\delta_{\bar{0}\bar{0}\bar{1}\bar{2}\bar{3}\bar{7}}]}\nonumber\\
&\times\frac{\Gamma[z_{12}+\delta_{01237}+2,z_{\bar{1}\bar{2}}+\delta_{\bar{0}\bar{2}\bar{3}\bar{7}}-1]}
{\Gamma[\delta_{7}+1,z_{12}+\delta_{\bar{4}56}+1]}
\label{template-pl1-13}\,.
\end{align}

\subsection*{Template Integrals of PL2}

The template integrals of
 $I^\mathrm{PL2}=J^\mathrm{PL2}_{1,1,1,1,1,1,1}$
defined in Eq.~\eqref{pl-2}
are
\begin{align}
\mathcal{T}^{(2)}=&
\int\mathrm{d}z_1\mathrm{d}z_2
\frac{\Gamma[\delta_{067},z_{\bar{1}}+\delta_{0156}+1,z_{\bar{1}},z_{1}+\delta_{\bar{0}\bar{3}\bar{4}\bar{5}}-1,z_{1}+\delta_{\bar{0}\bar{5}\bar{6}},z_{\bar{1}\bar{2}}+\delta_{001567},z_{\bar{2}},z_{2}+\delta_{6}+1]}
{(m_t^2)^{\delta_{001267}}S^{2+\delta_{34}}T^{1+\delta_{5}}
\Gamma[\delta_{1}+1,\delta_{2}+1,\delta_{\bar{0}\bar{3}\bar{4}\bar{5}}-1,\delta_{6}+1,\delta_{7}+1,z_{\bar{1}}+\delta_{0015667}+1]}\nonumber\\
&\times\frac{\Gamma[z_{12}+\delta_{2\bar{5}},z_{12}+\delta_{\bar{0}\bar{3}\bar{5}}]}
{\Gamma[z_{1}+\delta_{\bar{0}\bar{3}\bar{5}},z_{12}+\delta_{\bar{0}\bar{5}}+1]}
\label{template-pl2-2}\\
\mathcal{T}^{(3)}=&
\int\mathrm{d}z_1
\frac{\Gamma[\delta_{\bar{0}\bar{4}\bar{5}},\delta_{\bar{0}\bar{5}\bar{6}},\delta_{001567},z_{\bar{1}}+\delta_{0126}+1,z_{\bar{1}},z_{1}+\delta_{\bar{0}\bar{2}\bar{3}\bar{4}}-1,z_{1}+\delta_{\bar{2}5},z_{1}+\delta_{0\bar{2}567}]}
{(m_t^2)^{\delta_{001567}}S^{2+\delta_{34}}T^{1+\delta_{2}}
\Gamma[\delta_{1}+1,\delta_{\bar{0}\bar{2}\bar{3}\bar{4}}-1,\delta_{\bar{0}\bar{5}}+1,\delta_{5}+1,\delta_{7}+1,\delta_{0015667}+1,z_{1}+\delta_{\bar{0}\bar{2}\bar{4}}]}\label{template-pl2-3}\\
\mathcal{T}^{(4)}=&
\int\mathrm{d}z_1\mathrm{d}z_2
\frac{\Gamma[\delta_{047},z_{\bar{1}}+\delta_{0345}+1,z_{\bar{1}},z_{1}+\delta_{\bar{0}\bar{4}\bar{5}},z_{1}+\delta_{\bar{0}\bar{1}\bar{5}\bar{6}}-1,z_{\bar{1}\bar{2}}+\delta_{003457},z_{\bar{2}},z_{2}+\delta_{4}+1]}
{(m_t^2)^{\delta_{002347}}S^{2+\delta_{16}}T^{1+\delta_{5}}
\Gamma[\delta_{2}+1,\delta_{3}+1,\delta_{4}+1,\delta_{\bar{0}\bar{1}\bar{5}\bar{6}}-1,\delta_{7}+1,z_{\bar{1}}+\delta_{0034457}+1]}\nonumber\\
&\times\frac{\Gamma[z_{12}+\delta_{\bar{0}\bar{1}\bar{5}},z_{12}+\delta_{2\bar{5}}]}
{\Gamma[z_{1}+\delta_{\bar{0}\bar{1}\bar{5}},z_{12}+\delta_{\bar{0}\bar{5}}+1]}
\label{template-pl2-4}\\
\mathcal{T}^{(5)}=&
\int\mathrm{d}z_1
\frac{\Gamma[\delta_{\bar{0}\bar{4}\bar{5}},\delta_{\bar{0}\bar{5}\bar{6}},\delta_{003457},z_{\bar{1}}+\delta_{0234}+1,z_{\bar{1}},z_{1}+\delta_{\bar{2}5},z_{1}+\delta_{\bar{0}\bar{1}\bar{2}\bar{6}}-1,z_{1}+\delta_{0\bar{2}457}]}
{(m_t^2)^{\delta_{003457}}S^{2+\delta_{16}}T^{1+\delta_{2}}
\Gamma[\delta_{3}+1,\delta_{\bar{0}\bar{5}}+1,\delta_{5}+1,\delta_{\bar{0}\bar{1}\bar{2}\bar{6}}-1,\delta_{7}+1,\delta_{0034457}+1,z_{1}+\delta_{\bar{0}\bar{2}\bar{6}}]}\label{template-pl2-5}\\
\mathcal{T}^{(6)}=&
\int\mathrm{d}z_1\mathrm{d}z_2
\frac{\Gamma[\delta_{012},z_{\bar{1}},z_{1}+\delta_{2}+1,z_{1}+\delta_{5}+1,z_{\bar{1}}+\delta_{\bar{0}\bar{5}\bar{6}\bar{7}}-1,z_{\bar{1}\bar{2}}+\delta_{\bar{0}1\bar{3}\bar{4}\bar{5}\bar{6}\bar{7}}-2,z_{\bar{2}},z_{2}+\delta_{6}+1]}
{(m_t^2)^{\delta_{012}}S^{3+z_{1}+\delta_{034567}}T^{z_{\bar{1}}}
\Gamma[\delta_{1}+1,\delta_{2}+1,\delta_{4}+1,\delta_{5}+1,\delta_{6}+1,\delta_{\bar{0}\bar{0}\bar{4}\bar{5}\bar{6}\bar{7}}]}\nonumber\\
&\times\frac{\Gamma[z_{12}+\delta_{04567}+2,z_{\bar{2}}+\delta_{\bar{0}\bar{4}\bar{5}\bar{6}}-1]}
{\Gamma[\delta_{7}+1,z_{\bar{2}}+\delta_{\bar{0}12\bar{3}\bar{4}\bar{5}\bar{6}\bar{7}}-1]}
\label{template-pl2-6}\\
\mathcal{T}^{(7)}=&
\int\mathrm{d}z_1\mathrm{d}z_2
\frac{\Gamma[\delta_{023},z_{\bar{1}},z_{1}+\delta_{2}+1,z_{1}+\delta_{5}+1,z_{\bar{1}}+\delta_{\bar{0}\bar{4}\bar{5}\bar{7}}-1,z_{\bar{1}\bar{2}}+\delta_{\bar{0}\bar{1}3\bar{4}\bar{5}\bar{6}\bar{7}}-2,z_{\bar{2}},z_{2}+\delta_{4}+1]}
{(m_t^2)^{\delta_{023}}S^{3+z_{1}+\delta_{014567}}T^{z_{\bar{1}}}
\Gamma[\delta_{2}+1,\delta_{3}+1,\delta_{4}+1,\delta_{5}+1,\delta_{6}+1,\delta_{\bar{0}\bar{0}\bar{4}\bar{5}\bar{6}\bar{7}}]}\nonumber\\
&\times\frac{\Gamma[z_{12}+\delta_{04567}+2,z_{\bar{2}}+\delta_{\bar{0}\bar{4}\bar{5}\bar{6}}-1]}
{\Gamma[\delta_{7}+1,z_{\bar{2}}+\delta_{\bar{0}\bar{1}23\bar{4}\bar{5}\bar{6}\bar{7}}-1]}
\label{template-pl2-7}\\
\mathcal{T}^{(8)}=&
\int\mathrm{d}z_1
\frac{\Gamma[\delta_{07}-1,z_{\bar{1}},z_{1}+\delta_{2}+1,z_{1}+\delta_{5}+1,z_{1}+\delta_{0123456}+4,z_{\bar{1}}+\delta_{\bar{0}\bar{2}\bar{3}\bar{4}\bar{5}}-2,z_{\bar{1}}+\delta_{\bar{0}\bar{1}\bar{2}\bar{5}\bar{6}}-2]}
{(m_t^2)^{-1+\delta_{07}}S^{4+z_{1}+\delta_{0123456}}T^{z_{\bar{1}}}
\Gamma[\delta_{2}+1,\delta_{34}+2,\delta_{5}+1,\delta_{\bar{0}\bar{0}\bar{1}\bar{2}\bar{3}\bar{4}\bar{5}\bar{6}}-2,\delta_{16}+2,\delta_{7}+1]}\label{template-pl2-8}\,.
\end{align}

%%%%%%%%%%%%%%%%%%%%%%%%%%%%%%%%%%%%%%%%%%%%%%%%%%%%%%
\section{A Corollary of the Generalized Barnes Lemma}
\label{app:mb}

The generalized Barnes lemma~\eqref{solmb2-10}
\begin{align}
&
\int_{C}^{}
\frac{dz}{2\pi i}
\frac{\Gamma [a_1-z,a_2-z,b_1+z,b_2+z,b_3+z]}
{\Gamma (c+z)}
\nonumber\\
&=
\frac{\Gamma [a_1+b_1,a_2+b_1,a_1+b_2,a_2+b_2,a_1+b_3,a_2+b_3]}
{\Gamma [a_{12}+b_{13},a_{12}+b_{23},-b_3+c]}
\, _3F_2
\left( 
\begin{array}[]{c}
	a_1+b_3,a_2+b_3,a_{12}+b_{123}-c\\
	a_{12}+b_{13},a_{12}+b_{23}
\end{array}
;1
\right)
\, ,
\label{eb1}
\end{align}
contains the generalized hypergeometric function $_3F_2$.
We try to reduce $_3F_2$ to
a product of $\Gamma$-functions
on a case-by-case basis,
using the relations given in Refs.~\cite{hypgeo1,hypgeo2}.
In this appendix
we present a formula which is, 
to the knowledge of the author,
not published.
It yields a useful corollary of the generalized Barnres lemma~\eqref{solmb2-10}.

Consider the case where 
\begin{align}
a_1=n_1-b_1, a_2=n_2-b_2, \quad n_1, n_2\in \mathbb{N},
\label{int1}
\end{align}
where $\mathbb{N}$ is the set of positive integers.
Note that $n_1\not =0, n_2\not =0$
because otherwise the left pole and the right pole merge.

Substituting Eq.~\eqref{int1} into Eq.~\eqref{eb1},
the arguments of $_3F_2$ become
\begin{align}
	_3F_2\left( 
	\begin{array}[]{c}
		-b_1+b_3+n_1,-b_2+b_3+n_2,b_3-c+n_1+n_2\\
		-b_1+b_3+n_1+n_2,-b_2+b_3+n_1+n_2
	\end{array}
	;1
	\right)
	\, .
	\label{}
\end{align}
and here we express this type of $_3F_2$
in terms of the $\Gamma$-function only.
To this end, consider
\begin{align}
\mathcal{I}&=\, _3F_2\left( 
\begin{array}[]{c}
	x_1,x_2,x_3\\
	x_1+n_1,x_2+n_2
\end{array}
;1
\right)
\nonumber\\
&=
\sum_{m=0}^{\infty}\frac{1}{m!}
\frac{\Gamma [x_1+m,x_2+m,x_3+m,x_1+n_1,x_2+n_2]}
{\Gamma [x_1+m+n_1,x_2+m+n_2,x_3,x_1,x_2]}
\nonumber\\
&=
\sum_{m=0}^{\infty}\frac{1}{m!}
\left( \prod_{i_1=0}^{n_1-1} \frac{1}{x_1+m+i_1} \right)
\left( \prod_{i_2=0}^{n_2-1} \frac{1}{x_2+m+i_2} \right)
\frac{\Gamma [x_3+m,x_1+n_1,x_2+n_2]}
{\Gamma [x_3,x_1,x_2]}
\, ,
\label{}
\end{align}
where $x_1,x_2,x_3$ can contain $n_1,n_2$.
The products are resolved by the partial fraction decomposition 
\begin{align}
\prod_{n=0}^j \frac{1}{x+n}
=\sum_{n=0}^{j}\frac{(-1)^n}{n!(j-n)!} \frac{1}{x+n}
\end{align}
and we obtain
\begin{align}
\mathcal{I}&=
\sum_{m=0}^{\infty}\frac{1}{m!}
\sum_{i_1=0}^{n_1-1}
\sum_{i_2=0}^{n_2-1}
C_{i_1,i_2}
\frac{1}{(x_1+m+i_1)(x_2+m+i_2)}
\frac{\Gamma [x_3+m,x_1+n_1,x_2+n_2]}
{\Gamma [x_3,x_1,x_2]}
\\
&C_{i_1,i_2}=
\frac{(-1)^{i_1+i_2}}{i_1!i_2!(n_1-i_1-1)!(n_2-i_2-1)!}
\, .
\label{coef}
\end{align}
We can apply the partial fraction decomposition further and obtain
\begin{align}
\mathcal{I}&=
\sum_{m=0}^{\infty}\frac{1}{m!}
\sum_{i_1=0}^{n_1-1}
\sum_{i_2=0}^{n_2-1}
C_{i_1,i_2}
\frac{\frac{1}{x_1+m+i_1}-\frac{1}{x_2+m+i_2}}{x_2-x_1+i_2-i_1}
\frac{\Gamma [x_3+m,x_1+n_1,x_2+n_2]}
{\Gamma [x_3,x_1,x_2]}
\, .
\label{int7}
\end{align}
Here the equation in the case of $x_1-x_2\not\in\mathbb{Z}$
is shown, but the case of $x_1-x_2\in\mathbb{Z}$
is similar.
It is also possible to set $x_2\to x_1+n, n\in\mathbb{Z}$
at the end.
The infinite sum of $m$ in Eq.~\eqref{int7} is now possible,
and we obtain
\begin{align}
\mathcal{I}&=
\sum_{i_1=0}^{n_1-1}
\sum_{i_2=0}^{n_2-1}
\frac{C_{i_1,i_2}\Gamma [x_1+n_1,x_2+n_2]}
{(x_2-x_1+i_2-i_1) \Gamma [x_1,x_2]}
\left\{ 
\frac{\Gamma [x_1+i_1,1-x_3]}
{\Gamma [1+x_1+i_1-x_3]}
-\frac{\Gamma [x_2+i_2,1-x_3]}
{\Gamma [1+x_2+i_2-x_3]}
\right\}
\, .
\label{int8}
\end{align}

Substituting the replacement 
$x_1\to -b_2+b_3+n_2, x_2\to -b_1+b_3+n_1, x_3\to b_3-c+n_1+n_2$
into Eq.~\eqref{int8},
we obtain 
\begin{align}
&
\int_{C}^{}
\frac{dz}{2\pi i}
\frac{\Gamma [n_1-b_1-z,n_2-b_2-z,b_1+z,b_2+z,b_3+z]}
{\Gamma (c+z)}
\nonumber\\
&=
\sum_{i_1=0}^{n_1-1}
\sum_{i_2=0}^{n_2-1}
\frac{(-1)^{i_1+i_2}}{b_1-b_2-n_1+n_2+i_1-i_2}
\left\{ 
\frac{\Gamma [-b_1+b_3+n_1+i_2]}
{\Gamma [1-b_1+c-n_2+i_2]}
-\frac{\Gamma [-b_2+b_3+n_2+i_1]}
{\Gamma [1-b_2+c-n_1+i_1]}
\right\}
\nonumber\\
&\qquad \times
\frac{\Gamma [n_1,n_2,-b_1+b_2+n_1,b_1-b_2+n_2,1-b_3+c-n_1-n_2]}
{\Gamma [i_1+1,i_2+1,n_1-i_1,n_2-i_2,-b_3+c]}
\, .
\label{int9}
\end{align}
In this expression, there are only finite sums,
so they are evaluated in a straightforward way.
Among the six variables in Eq.~\eqref{int9},
two of them ($n_1,n_2$) should be positive integers,
but the other four ($b_1,b_2,b_3,c$) can take any value,
provided the left poles and right poles do not merge.

Once the integral is expressed in terms of the $\Gamma$-function,
we can easily compute derivatives of,
or analytically continue the result.

\bibliographystyle{JHEP}

\end{document}